\documentclass[aps,prb,twocolumn,groupedaddress]{revtex4}
\usepackage{amsmath,amssymb,bm,graphicx,dsfont,color}
\usepackage[latin9]{inputenc}
\usepackage{amsmath}
\usepackage{dsfont}
\usepackage{amstext}

\usepackage{amssymb}
\usepackage{amsbsy}
\usepackage{amsthm}
\usepackage{epsfig}
\usepackage{graphicx}
\usepackage{bbm}
\usepackage{hyperref}
\usepackage{color}
\begin{document}
\title{Physics of three dimensional bosonic topological insulators: Surface Deconfined Criticality and Quantized Magnetoelectric Effect.}
\author{Ashvin Vishwanath}
\affiliation{Department of Physics, University of California, Berkeley, CA 94720, USA}
\affiliation{Materials Science Division, Lawrence Berkeley National Laboratories, Berkeley, CA 94720}
\author{T. Senthil}
\affiliation{Department of Physics, Massachusetts Institute of Technology,
Cambridge, MA 02139, USA}
\date{\today}
\begin{abstract}
We discuss physical properties of `integer' topological phases of bosons in D=3+1 dimensions, protected by internal symmetries like time reversal and/or charge conservation. These phases  invoke interactions in a fundamental way but do {\em not} possess topological order and are bosonic analogs of free fermion topological insulators and superconductors.  While a formal cohomology based  classification of such states was recently discovered, their  physical properties remain mysterious. Here we develop a field theoretic description of several of these states and show that they possess unusual surface states, which if gapped,  must either break the underlying symmetry, or develop topological order. In the latter case, symmetries are implemented in a way that is forbidden in a  strictly two dimensional theory. While this is the usual fate of the surface states, exotic gapless states can also be realized. For example, tuning parameters can naturally lead to a deconfined quantum critical point or, in other situations,  a fully symmetric vortex metal phase. We discuss cases where the topological phases are characterized by quantized magnetoelectric response $\theta$, which, somewhat surprisingly, is an odd multiple of $2\pi$. Two different surface theories are shown to capture these phenomena - the first is a nonlinear sigma model with a topological term. The second invokes vortices on the surface that transform under a {\em projective} representation of the symmetry group. A bulk field theory consistent with these properties is identified, which is a multicomponent $BF$ theory supplemented, crucially, with a topological term. Bulk sigma model field theories of these phases are also provided.
A possible topological phase characterized by the thermal analog of the magnetoelectric effect is also discussed.

\end{abstract}

\maketitle

\tableofcontents
\section{Introduction}

Following the discovery of topological insulators\cite{TIs}, intense theoretical efforts have resulted in a good understanding of topological phases of free fermions, including a complete classification of such phases that are stable to disorder\cite{KitaevRyu} . In these phases the bulk appears, to local probes, as a rather conventional gapped state.   The surface however is gapless unless one of the symmetries protecting the phase is broken.

In contrast our understanding of topological phases of interacting particles is much less complete. The fractional quantum Hall effect has inspired much work on 
phases with topological order\cite{Wenbook}. These states have the remarkable property that the degeneracy of their ground states depends on the topology of the space on which they are defined. While they are often associated with edge states, additionally, excitations with exotic statistics occur in the bulk. Moreover, these states are characterized by a topological entanglement entropy, implying long range quantum entanglement in the ground state. It seems appropriate to exclude such phases in a minimal generalization of topological insulators to interacting systems.

We define a  Short Range Entangled ( SRE) state \footnote{This terminology differs slightly from that of Chen-Gu-Liu-Wen\cite{chencoho2011}, who require a state to also be non-chiral to be short range entangled.} as a gapped state with a unique ground state on all closed manifolds i.e. no topological order. In the presence of interactions, do new  SRE phases appear that share the same symmetry, but differ at the level of topology? A possible distinction, for example, is the presence of protected states at the boundaries.

In this paper we study the physics of such phases in systems of interacting bosons.   For bosons the non-interacting limit is a simple condensate so that interactions are necessary to stabilize gapped phases.  Thus we necessarily need to free ourselves from the crutch of free fermion Hamiltonians and band topology upon which most current discussions of topological insulators are based. Studying bosonic generaizations of topological insulators is potentially a useful step toward the harder problem of interaction dominated fermionic topological insulators.

Several examples of interacting boson systems exist. A very natural realization of a strongly interacting boson system is a quantum magnet made up of localized quantum spins on a lattice. In that context the phases we are interested in may be dubbed ``Topological Paramagnets".  They have a bulk gap, no bulk topological order or fractional quantum numbers, but have protected surface states all in close analogy to electronic topological insulators. The topological paramagnet should be distinguished from a more familiar (and more exotic) paramagnet - the quantum spin liquid state - which has been extensively discussed in the literature.  Quantum spin liquid states either have bulk topological order or gapless excitations. Other realizations of strongly interacting bosons are (of course) provided by ultracold atoms in optical lattices.

A famous example of a topological paramagnet already exists: the spin-$1$ Haldane (or AKLT) chain. This has a bulk gap, no fractionalization, but has dangling spin-$1/2$ moments at the edge that are protected by symmetry\cite{Auerbach1994}.
Using a powerful matrix product representation of gapped states\cite{Fidkowski2011,Turner2011,Chen2011a,Cirac}  topological phases in one dimension are completely classified by the second group cohomology of symmetry group G. In 2D and 3D such rigorous results are not available. However, it was pointed out by Kitaev that 2D SRE phases of bosons with chiral edge states  are nevertheless possible \cite{kitaevunpub}. Recently, Chen, Liu, Gu and Wen\cite{chencoho2011}  have proposed that SRE topological phases of bosons protected by symmetry, are captured by the higher dimensional cohomology groups.  While this provides an efficient mathematical scheme to enumerate phases, unfortunately their properties are not transparently obtained.

Progress on clarifying the physics of these states was subsequently made by several authors.  Levin and Gu\cite{LevinGu} explicitly studied a specific example in 2D protected by Z$_2$ symmetry.  Later, 2D SRE topological phases were described using a simpler Chern-Simons approach\cite{luav2012}, which provides a field theory and explicit edge theories of these states. It was found that bosons with a conserved global U(1) but no other symmetries can display an `integer quantum Hall state' with quantized Hall conductance predicted to be  even integer multiples of $q^2/h$, i.e. $\sigma_{xy}=2n \,q^2/h$ where $q$ is the elementary charge of the bosons\cite{luav2012}. A simple physical realization of such a phase of two component bosons in the lowest Landau level has been provided\cite{tsml}. Other physical interpretations have also appeared \cite{WenSU(2),ChenWen}. Thus, defining properties of several `Symmetry Protected Topological' (SPT) phases in 1D and 2D are now understood. However in 3D, though the cohomology classification predicts various SPT phases, including bosonic generalizations of topological insulators, their physical properties have not thus far been elucidated. This will be our primary task.  Potentially, our approach could also be used to classify 3D bosonic SPT phases, which we leave to future work. We note however that for all symmetry classes that we considered both with time reversal ($Z_2^T,\,U(1)\times Z_2^T,\,U(1)\rtimes Z_2^T$) and without ($U(1)\rtimes Z_2$) we identify candidate phases that exhaust those predicted by the cohomology classification\cite{chencoho2011}. Further, we identify a possible topological phase that appears to be outside the classification of Ref.\onlinecite{chencoho2011}.

 Much of our discussion will focus on a theory of the novel surface states of these three dimensional bosonic topological insulators. We construct effective ``Landau-Ginzburg"  field theories of these surface states. We also identify bulk field theories that correctly yield the proposed surface theory.  A key feature of this surface theory is that it does not admit a trivial gapped symmetry preserving surface phase. The surface either spontaneously breaks symmetry, or if gapped, develops surface topological order (even though there is no bulk topological order).   Other more exotic symmetry preserving states with gapless excitations are also possible. In all these cases the defining global symmetries are implemented  in a way not allowed in strictly two dimensional systems.
As a specific example - consider insulating bosonic phases with the symmetries of a topological insulator, i.e. charge conservation  and time reversal symmetry (formally denoted as $U(1)\rtimes Z_2^T$).  Then, the surface could break a symmetry, e.g. by forming a surface superfluid. Alternately, the surface can remain insulating, while breaking time reversal symmetry. The bulk is assumed to retain the symmetry. Then, these surface ordered phases can reflect their special origin by exhibiting features that are forbidden in purely two dimensional phases. For example, the vortices of the surface superfluid above will be shown, in a precise sense, to be fermionic 
Furthermore, the surface insulator with broken time reversal symmetry will be shown to have a Hall conductance of $\sigma_{xy}=\pm 1$ (in units of $q^2/h$), in contrast to the 2D integer Quantum Hall phases of bosons which are only allowed to even integer Hall conductance.

Equivalently, the surface Quantum Hall effect may be considered a quantized three dimensional response, the magneto-electric polarizability $\theta$. Recall, in the context of  free fermion topological insulators with broken time reversal on the surface that gaps the surface states, have a quantized magneto-electric effect appears\cite{QiHughesZhang} that is captured by the topological theta term:
\begin{equation}
{\mathcal L}_\theta = \frac{\theta}{4\pi^2} \vec E. \vec B
\label{Ltopo}
\end{equation}
(we have set $h=c=e=1$ and $\vec E,\,\vec B$ are applied electric and magnetic fields).  For fermion topological   insulators $\theta=\pi$ $({\rm mod} 2\pi)$, corresponding to a half integer Hall effect on the surface. The $2\pi$ ambiguity in $\theta$ corresponds to the fact that one may deposit a fermionic integer Quantum Hall layer on the surface\cite{emvtheta}.

Here,  for bosonic topological insulators, $\theta$ is {\em only} defined modulo $4\pi$, and the topological phase corresponds to $\theta=2\pi$. This implies, for example, the domain wall between opposite time reversal symmetry breaking regions on the surface induces a protected mode, the  edge state corresponding to the $\sigma_{xy}=2$ quantized Hall effect of bosons. The $4\pi$ ambiguity in $\theta$ corresponds to the fact that one may deposit an integer Quantum Hall layer of bosons on the surface, which must have an {\em even} integer Hall conductance.

A symmetry preserving surface state may be accessed from the superfluid by condensing vortices which transform trivially under the symmetry. However as the vortex of the surface superfluid has fermionic statistics it cannot condense. This precludes the possibility of a trivial gapped surface insulator. The fermionic vortices can of course pair and condense. However as is well known\cite{SenthilFisher} paired vortex condensation leads to a two dimensional state with topological order (described in the present context by a deconfined $Z_2$ gauge theory). This surface topological order will be shown to realize symmetry in a manner not allowed in strictly two dimensional systems.

Exotic  gapless surface states that preserve all symmetries are also conceivable. For example, the gapless surface state may intuitively be viewed as a Quantum Hall state that fluctuates between $\sigma_{xy}=\pm 1$.  The theory of such a state is constructed using a network model that captures the quantum phase transition between distinct integer quantum Hall states of bosons . The same approach, when applied to fermionic topological insulators correctly yields the single Dirac cone surface state.  The field theory thus obtained of the bosonic model poised at the transition, naturally leads to the required surface theory, which is closely related to the Deconfined Quantum Critical theory\cite{deccp}, previously proposed in the context of frustrated quantum magnets .

{\em 3D Field Theory:} The general arguments above will be shown to be consistent with the following $d=3$ field theory:
\begin{equation}
2\pi {\mathcal L}_{3D} = \sum_I\epsilon^{\mu\nu\lambda\sigma}B^I_{\mu\nu}\partial_\lambda a^I_\sigma + \Theta \sum_{I,J}\frac{K_{IJ}}{4\pi}  \epsilon^{\mu\nu\lambda\sigma}\partial_\mu a^I_\nu \partial_\lambda a^J_\sigma
\label{BFOverview}
\end{equation}
where the index $I$ refers to boson species, and bosons four currents are represented by $j^{\mu I} = \frac{1}{2\pi}\epsilon^{\mu\nu\lambda\sigma}\partial_\nu B^I_{\lambda\sigma}$ while the curl of $a^I$ represents the vortex lines. The first `BF' term\cite{BFreview} just represents the $2\pi$ phase factor of taking a particle around its vortex. The key topological properties however are determined by the second term, which attaches quantum numbers to vortices . To avoid topological order at the surface, and ensure bosonic excitations, ${\rm det} K=1$ and  diagonal entries are even integers. In most cases we will take two species with $K=\sigma_x$.
Here $\Theta \rightarrow \Theta+2\pi$ is assumed to lead to an equivalent theory that only differs in details of surface termination.
Furthermore, time reversal symmetry constrains $\Theta=0,\pi$, the latter being the topological phase.  Note, $\Theta$ for the internal gauge fields $a^I$ should be distinguished from the $\theta$ for the external electromagnetic field we discussed above. Coupling to an external electromagnetic field allows one to obtain the quantized magneto electric effect discussed above. Related theories have appeared in the context of three dimensional topologically ordered phases and superconductors \cite{OganesyanSondhi} where only the first $BF$ term in Eqn.\ref{BFOverview} appears with a different coefficient. On the other hand the field theory discussed in Ref. \onlinecite{WalkerWang} retains only the second term, which leads to gapless excitations in the bulk, as noted in Ref. \onlinecite{SteveSimonetal}, which differ from the gapped phases of interest here. Thus, it will be important to combine both these terms.

Recently, it was proposed that free {\em fermion} topological insulators  are captured by similar theories, \cite{ChoMoore} with a single component field and the first term of Eqn. \ref{BFOverview} along with coupling to the external field. The surface states in such theories were argued to be bosonized Dirac fermions\cite{Aratyn,ChoMoore}. While this is an intriguing idea, we point out  certain problems (see Appendix \ref{appendixBFsurface}) with the identification of a metallic surface in Ref. \onlinecite{ChoMoore}. Moreover, identification of a bulk fermionic operator is also problematic in this theory \onlinecite{Aratyn,ChoMoore}. One approach, taken in a recent paper\cite{RyuFradkin}, is to regard this as a partial theory, that provides a purely hydrodynamic description that excludes fermionic excitations altogether. Thus, finding a complete effective field theory description for 3D fermionic topological insulators remains an open problem.

We also display continuum field theoretic models in $D = 3+1$ dimension that realize some of the topological phases we describe. These are obtained as perturbations of non-linear sigma models in the presence of a topological theta term. The theta term has the effect of endowing topological defect configurations  of the continuum fields with non-trivial global quantum numbers.  We show the connection to the topological BF theory and to the theory of protected surface states.

{\em Possible Phases with Half Quantized Surface Thermal Hall effect:}  Interestingly,  our general approach and the theory in Eqn. \ref{BFOverview} predicts a new 3D SPT phase, protected by time reversal symmetry, that is not obtained within the cohomology classification of Chen et al.\cite{chencoho2011} Just as the quantized Hall effect of $d=2$ SRE  bosons immediately constrained the physics in 3D, an analogous argument for thermal Hall conductance can be made. In $d = 2$, this is quantized to 8 times the quantum of thermal conductance: $\kappa_{xy}/T = 8n\frac{\pi^2k_B^2}{3h}$, for a 2D SRE phase of bosons. A realization of $n=1$ is the Kitaev $E_8$ state\cite{kitaevunpub}, with 8 chiral bosons at the edge. Therefore, a three dimensional phase protected by time reversal symmetry can be conceived, on whose surface a domain wall between opposite T breaking regions will host 8 chiral boson modes and is described by Eqn. \ref{BFOverview},  with the 8 dimensional $K^{E_8}$ matrix, discussed in \onlinecite{luav2012} . This putative phase lies outside the cohomology classification of Chen et al.\cite{chencoho2011}, which reports a single non-trivial topological phase with this symmetry, for which a different candidate, with $K=\sigma_x$ is identified below.

\section{Overview}

As this paper is long and discusses bosonic SPT phases from several points of view it is helpful to provide an overview. We seek an effective ``Landau-Ginzburg" description of the surface of a 3D SPT phase. We employ various approaches, that, satisfyingly, lead to consistent conclusions.  Below we motivate these different approaches as preparation for the rest of the paper.

A key physical requirement of the surface theory is that there be no trivial gapped symmetry preserving surface phase. This feature is reminiscent of the Lieb-Schutlz-Mattis (LSM) theorem \cite{Auerbach1994} and its generalization\cite{Oshikawa,Hastings} to states of bosonic systems at a fractional filling on clean $2d$ lattices. Indeed in both cases either a symmetry must be broken or there is topological order, or more exotic long range entanglement possibly with protected gapless excitations.  There is however one important difference. For the surface states discussed in this paper, the trivial insulating phase does not exist {\em even} in the presence of disorder that breaks translation symmetry. They are protected by internal symmetries rather than by lattice translation symmetry (as is the case for LSM).  Nevertheless we will exploit insights from existing effective field theories of clean lattice bosons that build in the LSM restrictions to construct the desired surface theory of the bosonic topological insulator.

We begin with a physical discussion of the constraints on quantized electrical and thermal response in SRE 3D bosonic insulators, imposed by our knowledge of 2D SRE phases ({\bf Section} \ref{Sec:gentrnsprt}). In particular we argue that the quantized magnetoelectric coupling for bosonic topological insulators ($\theta=2\pi$) is double that of the fermionic case ($\theta=\pi$), which parallels the doubling of Hall conductance\cite{luav2012,tsml,ChenWen} for 2D integer states of bosons compared to fermions.

Next, we borrow an approach that is useful in classifying SPT phases of 2D bosons. There, the edge states are typically described by a 1D Luttinger liquid theory. What makes them special is that symmetry acts on these edge states in a way that is impossible to realize in a purely 1D system\cite{luav2012,WenSU(2),tsml,ChenWen}. This ensures that all perturbations that lead to a  gapped phase also break symmetry. By analogy, to discuss a 3D SPT phase we model the 2D surface by a conventional 2D  theory of bosons (e.g. an XY model). However, we identify symmetry operations which are forbidden in a conventional 2D system, by demanding the 2D system can never enter a trivial gapped phase.

\begin{figure}
\begin{centering}
\includegraphics[scale=0.20]{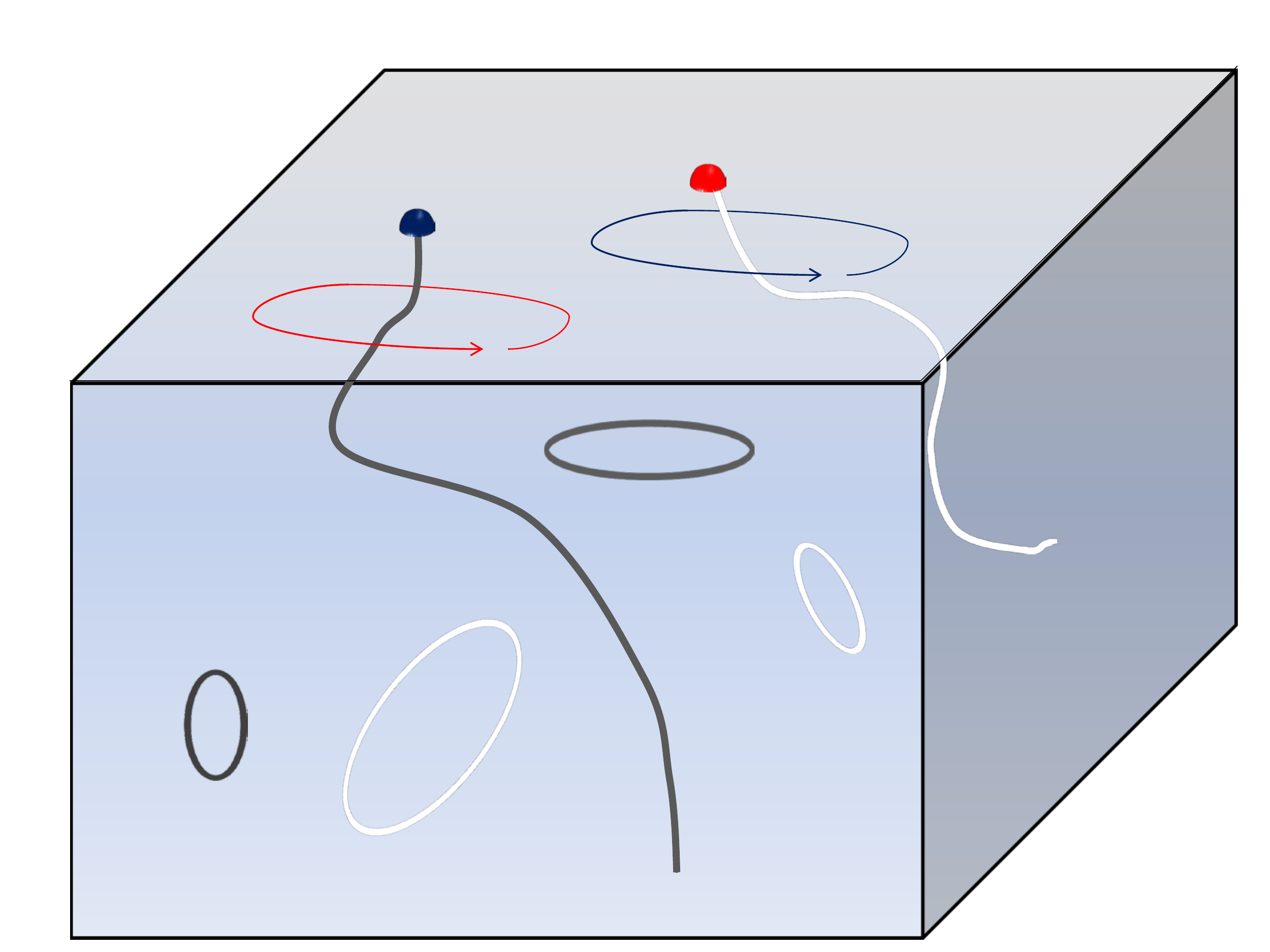}
\par\end{centering}
\caption{Schematic depiction of a 3D symmetry protected topological phase, with two conserved species of bosons (U(1)$\times$U(1)) and time reversal symmetry ($Z_2^T$). The bulk is insulating and corresponds to a condensate of vortices of both species (shown as black and white loops). In the topological phase, the vortex line of one species that ends on a surface carries half charge of the other species. Such a surface, it may be argued, does not have a trivial gapped phase, where the symmetries are preserved. Pictorially, the vortex lines may be viewed in 1D Haldane phases, with half charged end states. }
\label{Fig1}
\end{figure}

As a useful device consider an enlarged $U(1) \times U(1)$  global symmetry, along with time reversal symmetry $\mathcal T$  (technically $\left [ U(1)\times U(1)\right ] \rtimes Z_2^T$)  and eventually break it down to the symmetry of interest. Physically, this corresponds to separate conservation of two species of bosons Species 1,2. Consider the  surface of a 3D SPT phase, with a broken symmetry eg. a condensate of Species 1 bosons, that breaks the first U(1) symmetry. This superfluid must also be unusual - in that it cannot be connected to a fully symmetric insulator. Guidance from effective field theories of clean 2d lattice bosons at fractional filling suggests thinking in terms of the vortices of the superfluid. Since the insulating state is obtained by condensing vortices, we can ask what vortex properties provide the required obstruction. Unlike a particle, a vortex is a non-local object, and can transform projectively under the remaining $U(1)$ and $\mathcal T$ symmetry.  Vortices with projectively realized symmetries provide an obstruction to realizing a trivial insulator, and can describe the surface of an SPT phase. In this example, the projective transformation requires the two species of vortices ( $\psi_\pm$) carry half charge of Species 2, which are exchanged by time reversal symmetry. A minimal theory of these vortices is obtained by representing the density of the condensed Species 1 by the curl of a vector potential $N_1 = (\partial_x \alpha_y-\partial_y \alpha_x)/2\pi$, which couple minimally to the vortices.

\begin{eqnarray}
{\mathcal L}_{\rm edge2D} &=& \sum_{\sigma=\pm}|(\partial_\mu -i\alpha_\mu-\frac{i\sigma}{2}A_{2\mu})\psi_\sigma|^2  + \frac{\epsilon^{\mu\nu\lambda}}{2\pi} A_{1\mu}\partial_\nu \alpha_\lambda \nonumber \\&&+K (\partial_\mu \alpha_\nu -\partial_\nu\alpha_\mu)^2
\label{DECCPSurface}
\end{eqnarray}
Where we have inserted external electromagnetic fields $A_1, \,A_2$ that couple to the two conserved currents. To obtain an insulating surface we must condense the vortices, but this inevitably breaks symmetry. Since the vortices transform into one another under time reversal, one cannot condense one species and not the other. Condensing them both however implies the breaking the other U(1) symmetry, since the vortices carry charge of Species 2. A third option is to condense pairs of vortices, but this results in Z$_2$ topological order\cite{SenthilFisher}. This points to a schematic picture of the SPT phase as shown in Figure \ref{Fig1}. A bulk insulator may be regarded as a condensate of vortex loops. However, the special feature here is  that the condensed vortex lines carry a half charge at their ends when they intersect the surface. The vortex lines, viewed as one dimensional objects, are in a topological state analogous to the Haldane phase, and thus carry edge states.

Note that if a single bosonic vortex species was present at low energies, we could perform a duality back to the usual boson phase variables. However, having multiple vortex fields demands a dual description like the one above. In fact an identical theory appeared in the discussion of Deconfined Quantum Critical points\cite{deccp,lesik12a,Courtney}, where however the symmetries included both internal and spatial symmetries. Here all symmetries are internal and hence are restricted to the surface of a 3D topological phase.

Let us consider in more detail the insulating surface obtained on breaking time reversal symmetry. This is obtained by condensing  just one species of vortex say $\psi_{-}$, in Eqn.\ref {DECCPSurface} which forces $\alpha=\frac12 A_2$. When substituted into the action above this yields  the electromagnetic response $${\mathcal L}_{em} = \frac{1}{4\pi}A_{2\mu}\epsilon_{\mu\nu\lambda} \partial_\nu A_{1\lambda}$$. If the separate species are now identified with a single conserved charge, we can set $A_1=A_2=A$  which yields  $\sigma_{xy}=1$ on the surface, indicating a magneto electric response $\theta=2\pi$, as advertised.  In the absence of time reversal breaking, the surface may be assumed to fluctuate between $\sigma_{xy}=\pm1$. We provide an alternate derivation of the same surface theory Eqn.\ref{DECCPSurface} by modeling this by a  network model, poised at the transition point between two bosonic integer Quantum Hall states.  This is analogous to obtaining the surface state of the 3D fermionic topological insulators, the single Dirac cone, by a network model at the critical point between integer quantum Hall plateaus\cite{ludwigfisher} in a clean system. In both cases, the time reversal symmetry automatically tunes the system to criticality. For the bosonic SPT surface, resulting theory is an O(4)  non-linear sigma model with a topological term as in the Euclidean Lagrangian below:
\begin{eqnarray*}
{\mathcal L}'_{edge2D} = &&\frac1{2\kappa} {\rm Tr}(\partial_\mu g^\dagger \partial_\mu g)\\
&+&i\frac{\pi}{24\pi^2} \epsilon^{\mu\nu\lambda} {\rm Tr}[(g^\dagger\partial_\mu g )(g^\dagger\partial_\nu g)( g^\dagger\partial_\lambda g)]
\end{eqnarray*}
where the O(4) vector has been written in terms of an SU(2) matrix $g$. Reassuringly, this has been argued in Ref.  \onlinecite{tsmpaf2006}  to be equivalent to Eqn. \ref{DECCPSurface} once appropriate anisotropies are introduced.  These two descriptions of surface properties of SPT phases are discussed in {\bf Section} \ref{Sec:surface} along with the connection between them.

We also emphasize the properties of a gapped symmetric surface state obtained by condensing paired vortices. This state has surface topological order described by a deconfined $Z_2$ gauge theory (even though the bulk has no such order).  This state provides a particularly simple perspective on why a trivial gapped symmetric phase is forbidden at the surface. Indeed we will show that bosonic topological quasiparticles of this state carry fractional quantum numbers. Destroying the topological order by condensing one of these quasiparticles necessarily breaks a symmetry. Not surprisingly we show that the implementation of symmetry in this surface topological ordered state is distinct from what is allowed in strict 2d systems.

In the remaining sections bulk theories that are consistent with the surface descriptions above are discussed in {\bf Section} \ref{Sec:3D}. SPT phases in quantum magnets - the ones we dubbed Topological Paramagnets - protected by other symmetries such as time reversal ($Z_2^T$), or  time reversal along with one component of spin rotation  ($U(1)\times Z_2^T$) are described in {\bf Section} \ref{Sec:OtherSymm}. An example without time reversal symmetry ($U(1)\rtimes Z_2$) is discussed in the appendix. Finally, we discuss a new topological phase that is predicted by this approach, a 3D extension of Kitaev's E$_8$ state, along with miscellaneous comments in {\bf Section} \ref{Sec:E8}.

\section{Transport Properties of 3D bosonic topological insulators: General Constraints}
\label{Sec:gentrnsprt}
We begin by considering a system of interacting bosons in $d = 3$ space dimensions in the presence of time reversal and particle number conservation symmetries. Specifically let us consider the situation where the boson field $b$ carries charge $1$ under a global $U(1)$ symmetry and transforms as $b \rightarrow b$ under time reversal. The corresponding symmetry group is $U(1) \rtimes Z_2^T$. Assume it is in a gapped insulating phase (at least in the absence of any boundaries) and that there is a unique ground state on topologically non-trivial manifolds.  For any such insulator in 3D, the effective Lagrangian for an external EM field obtained by integrating out all the matter fields will take the form
\begin{equation}
{\cal L}_{eff} = {\cal L}_{Max} + {\cal L}_\theta
\end{equation}
The first term is the usual Maxwell term and the second is the `theta' term in Eqn. \ref{Ltopo}:

Several properties of the theta term are well known. First under time reversal, $\theta \rightarrow - \theta$. Next on closed manifolds, the integral of $\frac{1}{4\pi^2} \vec E. \vec B$ is quantized to be an integer so that the quantum theory
is periodic under $\theta \rightarrow \theta + 2\pi$. These two facts together imply that time reversal symmetric insulators have $\theta = n \pi$ with $n$ an integer. Trivial time-reversal symmetric insulators have $\theta = 0$ while free fermion topological insulators have $\theta = \pi$.

If we allow for a boundary to the vacuum and further assume that the boundary is gapped (if necessary by breaking time reversal symmetry), then the $\theta$ term leads to a surface Hall conductivity of $\frac{\theta}{2\pi}$.
To see this, assume a boundary (say at $z = 0$), $\theta = \theta(z)$ is zero for $z< 0$ and constant $\theta$ for $z >  0$. The action associated with the $\theta$ term
is
\begin{eqnarray}
S_\theta & = & \frac{1}{8\pi^2} \int d^3x\, dt \,\theta(z) \partial_\mu K^\mu \\
& = & -\frac{1}{8\pi^2} \int d^3x\,dt \, \frac{d\theta}{dz} K^z \\
& = & \frac{\theta}{8\pi^2} \int_{\partial B} d^2x\,dt \epsilon^{z\nu \lambda \kappa} A_\nu \partial_\lambda A_\kappa
\end{eqnarray}
Where $A$ is the external electromagnetic potential and $K^\mu = \epsilon^{\mu\nu\lambda\kappa} A_\nu \partial_\lambda A_\kappa$. This is a surface Chern-Simons term and leads to a Hall conductivity $\theta/2\pi$.

For fermion topological insulators $\theta = \pi$ so that the surface $\sigma_{xy} = \frac{1}{2}$. If we shift $\theta \rightarrow \pi + 2n\pi$, then the surface $\sigma_{xy} = (n + \frac{1}{2})$. This corresponds to simply depositing an ordinary integer quantum Hall state of fermions at the surface of this insulator -  hence this should not be regarded as a distinct bulk state so that the only non-trivial possibility is $\theta =  \pi$.

Now let us consider bosonic insulators. Again T-reversal and periodicity imply $\theta = n\pi$ and a surface $\sigma_{xy} = n/ 2$. A crucial observation is that now $\theta = 2\pi$ must be regarded as {\bf distinct} from $\theta = 0$. At $\theta = 2\pi$ the surface $\sigma_{xy} = 1$. But this cannot be obtained from the surface of the $\theta = 0$ insulator by depositing any 2d integer quantum Hall state of bosons. Recent work\cite{luav2012,tsml} has shown (see Ref. \onlinecite{tsml} for a simple argument) that 2d IQHE states of bosons necessarily have $\sigma_{xy}$ even. Thus the surface state of the $\theta = 2\pi$ boson insulator is not a trivial 2d state but rather requires the presence of the 3d bulk.

Therefore $\theta = 2\pi$ necessarily corresponds to a non-trivial 3d bosonic TI. $\theta = 4\pi$ is however trivial as then the surface state can be regarded as a 2d bosonic IQHE state. One may still obtain a 3D topological phase, but the topology is not manifest in the electromagnetic response.

We can sharpen and generalize this result. Under T-reversal as $\theta \rightarrow -\theta$, $n\pi \rightarrow - n\pi$. As the bulk state is T-reversal invariant we require that the surface state at $\theta = -n \pi$ be obtainable from the surface state at $\theta = +n \pi$ by depositing a 2d IQHE boson state. Let us characterize the surface state by both its electrical and thermal hall conductivities
$(\sigma_{xy}, \kappa_{xy})$. Under T-reversal both Hall conductivities change sign. The requirement above then means that $(2\sigma_{xy}, 2\kappa_{xy})$
must correspond to the allowed electrical/thermal Hall conductivity of a 2d boson IQHE state.

For $\theta = 2\pi$, it follows that $\sigma_{xy} = 1, \kappa_{xy} = 0$. It is thus ``half" of the elementary 2d boson IQHE state.

For $\theta = \pi$, $2\sigma_{xy} = 1$ and this is not allowed for the 2d bosonic IQHE. It follows therefore that 3d bosonic TIs with no ``intrinsic topological order" cannot have $\theta = \pi$. It is of course very easy to construct such states\cite{swingle3dfti,levin3dfti,macj3dfti} (or other states with fractional $\theta$) if we allow for fractionalization of the boson but that violates our original assumption.

A 2d IQHE state with $2\sigma_{xy} = 8, 2\kappa_{xy} = 8$ is allowed and is discussed by Kitaev. Thus a 3d boson TI with surface $\sigma_{xy} = 4, \kappa_{xy} = 4$ is allowed. Combining these two types of fundamental states generates the allowed thermal and electrical Hall responses on the surface.

Later on in the paper we will discuss how these results fit in with the formal classification of SPT and other short ranged entangled phases in 3d. For now we reiterate the crucial observation of this section: {\em A state with EM response of $\theta = 2\pi$ necessarily describes  a topological insulator of $T$-reversal symmetric bosons while $\theta = \pi$ requires the presence of ``intrinsic topological order"}.
In the next Section we study the properties of this $\theta = 2\pi$ boson topological insulator in detail.

\section{Surface Theory of 3D Bosonic SPT Phases}
\label{Sec:surface}
In this section we will derive the non-trivial surface theory of one example of a  3D bosonic SPT phase.  We will soon specialize to the symmetries of the topological insulator: charge conservation and time reversal symmetry ($U(1)\rtimes Z^T_2$), and exhibit a nontrivial topological phase in three dimensions, built purely of bosons. To begin with we will assume that there are two species of bosons whose numbers are separately conserved, and there is enlarged $\left(U(1) \times U(1)\right) \rtimes Z^T_2$. Later we will break this to just  $U(1) \rtimes Z^T_2$ symmetry by including inter-species boson mixing terms in the Hamiltonian. A similar construction \cite{luav2012} has proven to be very powerful in $d = 2$.  We consider two approaches:

The first approach exploits the fact that a bosonic SPT phase in $d$ dimensions has surface states that  correspond to a conventional theory of bosons in $d-1$ dimensions {\em except} in the way symmetries are implemented. For example, the edges of SPT phases of bosons in D=2+1 dimensions correspond to conventional 1D Luttinger liquids, except for their unusual symmetry transformations \cite{luav2012,tsml,WenSU(2)}. We therefore consider a 2D bosonic state to model the surface and assume the surface is a superfluid breaking one of the U(1) symmetries. Then, vortices of this condensate  may transform under a projective representation of the remaining symmetry group. In a projective representation, even the identity element of the symmetry group induces a phase rotation.  Hence local operators, that can be physically measured, must remain unchanged under the identity operation of the symmetry group, since this corresponds to `doing nothing'.  However, vortices, which are  non-local objects, can transform projectively. One may attempt to restore the U(1) symmetry by condensing vortices. However, the projective transformation ensures that when vortices condense, they  necessarily break another symmetry. In this way both the boson and vortex condensates lead to symmetry breaking, in line with our general expectation for the surface of a 3D SPT phase. It is important that vortices transform projectively, so that they cannot be screened by bosons to obtain a trivial representation of the symmetry group.   This is a generalization of the idea of quantum number fractionalization - for example a particle with half charge changes sign under the $2\pi$ phase rotation of bosons, implying a projective representation. Clearly, a half charge cannot be screened by any finite number of bosons. Projective representations were also  recently used to classify SPT phases in D=1+1where they correspond to the ends of gapped one dimensional topological phases. For example, the half-integer spin edge states of spin-1 Haldane chains furnishes a projective representation of the rotation group. This suggests a physical picture of a 3D SPT phase, in which the vortex line in the bulk is similar to a Haldane chain type gapped phase, which necessitates low energy states on the surface where the vortex ends.  In this section we specialize to the symmetries of the topological insulator. Then this procedure explicitly produces a topological phase characterized by quantized magnetoelectric effect $\theta=2\pi$.

The second approach will be to directly implement the property discussed above in Section \ref{Sec:gentrnsprt} that if the surface breaks $T$-reversal and is gapped then it has quantized Hall transport. If $T$-reversal is not broken a powerful approach to obtain the surface theory is to start with the theory of the quantum phase transition point between the two bosonic quantum Hall phases that correspond to the two $T$-broken surfaces. In the case of free fermion topological insulators, a similar reasoning leads to the single Dirac cone surface state that describes the transition between the $\sigma_{xy} = \pm \frac{1}{2}$ states on the surface. For free fermions the transition between these integer quantum Hall states is described by a Chalker-Coddington network model\cite{chalker}. For the bosonic problem of interest here we construct an analogous network model and show it leads to a sigma model with a topological term.

The results of these two approaches are readily seen to be connected. In both cases the field theories we obtain for the surface have appeared previously in the context of deconfined quantum criticality. We discuss the phase diagram of the surface states described by these field theories. When inter-species tunneling is included the vortices of the two species of bosons get confined to each other. The resulting single vortex no longer transforms projectively under the physical symmetries. However we argue that it is most conveniently viewed as a fermion. This precludes the possibility of obtaining a trivial insulating phase at the surface by condensing vortices.

\subsection{Surface States and Projective Vortices}

Consider a boson field at the surface with phase degree of freedom $\phi_1,\,b_1^\dagger = e^{i\phi_1}$. We assume the bulk is insulating and the surface is in the $x,\,y$ plane.

The surface theory could spontaneously break a global $U(1)$ symmetry of boson number conservation (a surface superfluid) or stay insulating. More precisely as the bulk is always assumed insulating, the vortex line loops have proliferated in the bulk. These vortex lines penetrate the surface at points, which may be viewed as point vortices of the two dimensional surface theory, since there is no vortex line tension in the insulating bulk. These point vortices are gapped when the surface is a superfluid. If instead they are condensed the surface will be insulating.
To describe vortices we go to a dual description\cite{DasguptaHalperin,FisherLee}, where we write the density and currents of the boson $b_1$ on the surface in terms of the field strengths of a gauge field $j_{1\mu}=\epsilon^{\mu\nu\lambda}\partial_\nu\alpha_{2\lambda}/2\pi$. In particular the density of bosons is: $n_1=(\partial_x \alpha_{2y}-\partial_y \alpha_{2x})/2\pi$ (the reason for the subscript 2 on $\alpha$ will soon be apparent). The boson insertion operators $e^{im\phi_1}$ correspond to monopole insertion operators, since they insert $2\pi m$ magnetic flux. Now, the vortices $\Psi_2$ are particles  that couple minimally to the gauge field $\alpha_2$. In general there will be multiple vortex species that transform into each other under the symmetry operation. Let us label them by $i$, so $[\Psi_2]_i=\psi_{2i}$. All these fields couple minimally to the gauge field.

Thus we have for the dual surface theory:
\begin{eqnarray}
{\mathcal L}_{\rm surf} &=& \sum_{i,\mu}|(\partial_{1\mu} -i\alpha_{2\mu})\psi_{2i}|^2 \nonumber \\&&+V(\Psi_{2i})+\frac{1}{2\kappa} f^2_{2\mu\nu}+\dots
\label{NCCP1vortex}\\
\end{eqnarray}
where
$f_{2\mu\nu}=(\partial_\mu\alpha_{2\nu}-\partial_{\nu}\alpha_{2\mu})$.

As argued above, one route to obtaining topological surface states is if the surface vortices  transform under a projective representation of the remaining symmetry. Vortices that transform projectively under a global symmetry are actually not at all unfamiliar. It describes the generic situation of two dimensional bosons on a lattice, say, at some commensurate filling\cite{Courtney,SachdevBalentsSengupta}. These projective vortices play a crucial role in the theory of deconfined quantum criticality.

We will only need to consider two component vortex fields, $\Psi_2=(\psi_{2+},\,\psi_{2-})$ for a variety of cases considered in this paper. The gauge invariant combination $\Psi_2^\dagger \sigma^+\Psi_2=\psi_{2+}^*\psi_{2-}=e^{i\phi_2}=b_2^{\dagger}$ then defines another bosonic field . Now, Eqn. \ref{NCCP1vortex} closely resembles the action for a deconfined quantum critical point (non-compact CP$_1$ theory with easy plane anisotropy) \cite{lesikav2004,deccp}. We will demonstrate how this emerges as the theory for the surface states, and also, in the next section, describe a three dimensional bulk theory that leads to this edge theory.

\subsubsection{Surface States of a Bosonic topological insulator: Symmetry U(1)$\rtimes\ Z^T_2$}
\label{Sec:bosonicTI}
These are the symmetries of the topological insulator, a conserved $U(1)$ charge and $Z_2^T$ time reversal symmetry.  The semi-direct product appears  so that the charge insertion operator $e^{i\phi}$ is invariant under time reversal, which involves both $\phi\rightarrow -\phi$ and $i\rightarrow -i$. Here we will construct the surface theory of a 3D topological phase with these symmetries.

Let us begin with an enlarged symmetry, two species of bosons that are separately conserved. Consider a condensate of one species $b_1$. Vortices in this condensate are created by the field $\Psi_2$. We need to specify the projective representation for the vortices $\Psi_2$ and the transformation of the bosons $b_1^\dagger$. The remaining symmetry group $U(1)\rtimes\ Z^T_2$ has a single projective representation (P1) which acts as follows. Under a $U(1)$ rotation by angle $\epsilon$, the fields $\psi_{2\pm} \rightarrow e^{\pm i\epsilon/2}\psi_{2\pm}$ and under time reversal: $Z_2^T$: $\psi_{2+} \rightarrow \psi^*_{2-}$ and $\psi_{2-} \rightarrow \psi^*_{2+}$. Or more compactly:

\begin{eqnarray}
\Psi_2 \rightarrow e^{i\frac\epsilon2 \sigma_z}\Psi_2 &&:U(1)\nonumber\\
\Psi_2 \rightarrow \sigma_x \Psi^*_2 &&:Z^T_2
\label{P1P1Vortex}
\end{eqnarray}
Here the $\sigma$  are the Pauli matrices in the standard representation. Thus the vortices carry charge $\pm 1/2$, of bosons of the other species. The time reversal symmetry that interchanges the two vortex fields ensures that the vortex charge is fixed exactly at half. It is impossible to `screen' this charge with regular integer charged  bosons.

These transformation laws of course determine how the boson operator $b^\dagger_2=e^{i\phi_2}$ and their density $n_2$ transforms. We also need to specify how the bosons $b^\dagger_1=e^{i\phi_1}$ and density $n_1$ transforms.  The symmetry transformations are:
\begin{eqnarray}
\phi_{1,2} & \rightarrow & \phi_{1,2} + \epsilon~~~~~~U(1) \nonumber\\
\phi_{1,2} & \rightarrow & -\phi_{1,2}~~~~~~~~~~~~~Z^T_2
\label{phiTransform}
\end{eqnarray}
The conjugate boson numbers therefore transform as
\begin{eqnarray}
n_{1,2} & \rightarrow & n_{1,2} ~~~~~~U(1)\nonumber \\
n_{1,2} & \rightarrow & n_{1,2}~~~~~~~~~~~~~Z^T_2
\label{nTransform}
\end{eqnarray}

A necessary compatibility check is that Eqn. \ref{NCCP1vortex} is invariant under the symmetry operation, which can be verified for these transformations. For example, time reversal symmetry is implemented via Eqn. \ref{P1P1Vortex} on the vortex fields which is compatible with $n_2$ (and hence $\alpha_2$) remaining invariant while $i\rightarrow -i$ under time reversal.   Moreover, since the bosons carry charge, the monopole insertion operators are forbidden.

 Condensing single vortices will then break symmetry as described below. This is best analyzed by assuming separate number conservation of each species of boson, $n_1,\,n_2$, in which case they can be coupled to an {\em external} gauge potential $A_{1},\,A_{2}$. Then the effective Lagrangian Eqn\ref{NCCP1vortex} reads:

\begin{eqnarray}
{\cal L}_{1} & = &
\sum_{s=\pm} \left|\left(\partial_\mu - i \alpha_{2\mu} -i\frac{sA_2}{2}\right)\psi_{2s}\right|^2 + .... \nonumber \\
&   & + \frac{1}{2\kappa_1} \left( \epsilon_{\mu\nu\lambda} \partial_\nu \alpha_{2\lambda} \right)^2 + \frac{1}{2\pi}A_{1\mu}\epsilon_{\mu\nu\lambda} \partial_\nu \alpha_{2\lambda}
\label{vortexaction2}
\end{eqnarray}
Here $s = \pm$.
The $\psi_{2\pm}$ are vortex fields of $b_1$ which carry half charge of boson species 2. The flux of the gauge field $\alpha_1$ is precisely the conserved density of species $1$; hence the last term in the above action where the external probe gauge field $A_1$ couples to this current. This action will need to be modified by including all symmetry allowed perturbations. We will do this and analyze the possible phases below, but as a preview, consider the effect of breaking time reversal symmetry by condensing just one species of vortex say $\psi_{2-}$. Anticipating a single charge, consider an external field that couples equally to the two charge densities $A_1=A_2=A$. The vortex condensate forces $\alpha_2=\frac12 A$, which when substituted into the action above yields  the electromagnetic response $\frac{1}{4\pi}A_{\mu}\epsilon_{\mu\nu\lambda} \partial_\nu A_{\lambda}$, which yields  $\sigma_{xy}=1$ on the surface, indicating a magneto electric response $\theta=2\pi$, as advertised.  We now change track and obtain the same surface theory from a very different point of view.

\subsection{Network Model Construction}
The general considerations of Section \ref{Sec:gentrnsprt} showed that a time reversal symmetric boson insulator with electromagnetic response characterized by $\theta = 2\pi$ is in a topological insulator phase. This key result relied on the observation that if ${\cal T}$ is broken at the surface to gap it out then such a state has a quantized electrical Hall conductivity $\sigma_{xy} = \pm 1$, and a thermal Hall conductivity $\kappa_{xy} = 0$.
What if T-reversal is not explicitly broken at the surface? The surface can then potentially be gapless. What is the nature of the resultant theory?
To construct this theory it is extremely instructive to learn from the example of the free fermion $T$-reversal symmetric topological insulator. In that case if $T$ is explicitly broken to gap out the surface, then we get $\sigma_{xy} = \pm \frac{1}{2}$. When $T$ is unbroken, it is possible to get a single massless Dirac cone which is exactly the low energy theory of the transition between two integer quantum hall plateaus of fermions in $d = 2$. Generically we can tune the chemical potential to move away from the Dirac point to get a Fermi surface which encloses the Dirac point.

Note that the quantized Hall conductance jumps by 1 across the integer quantum Hall plateau transition. When applied to the surface of the 3d fermionic topological insulator, the
transition connects the two possible $T$-breaking surface states that go into each other under time reversal. Thus each such surface state must be assigned $\sigma_{xy} = \pm \frac{1}{2}$ and the critical theory itself is time reversal invariant.  In contrast when applied to strictly 2d systems the network model describes a transition between plateaus with $\sigma_{xy} = 0$ and $\sigma_{xy} = 1$.

The familiar free fermion example gives us a crucial clue to construct the theory of the $T$-reversal symmetric surface state of the boson topological insulator. First construct the low energy theory of the $d = 2$ integer quantum Hall state of bosons as a potential candidate for the gapless surface state of the $3d$ topological insulator.  Then add perturbations allowed by symmetry to obtain the generic surface theory.
 Across the boson integer quantum Hall transition, $\sigma_{xy}$ jumps by 2. As for the fermionic example, when this transition is realized at the surface of the bosonic topological insulator, the two plateau states on ether side of the transition are related by $T$-reversal and must be assigned $\sigma_{xy} = \pm 1$ consistent with our earlier arguments.

With this motivation we now study the IQHE plateau transition of bosons in $d = 2$.

{\bf  IQHE quantum phase transition of bosons in $d = 2$}

\begin{figure}
\begin{centering}
\includegraphics[scale=0.3]{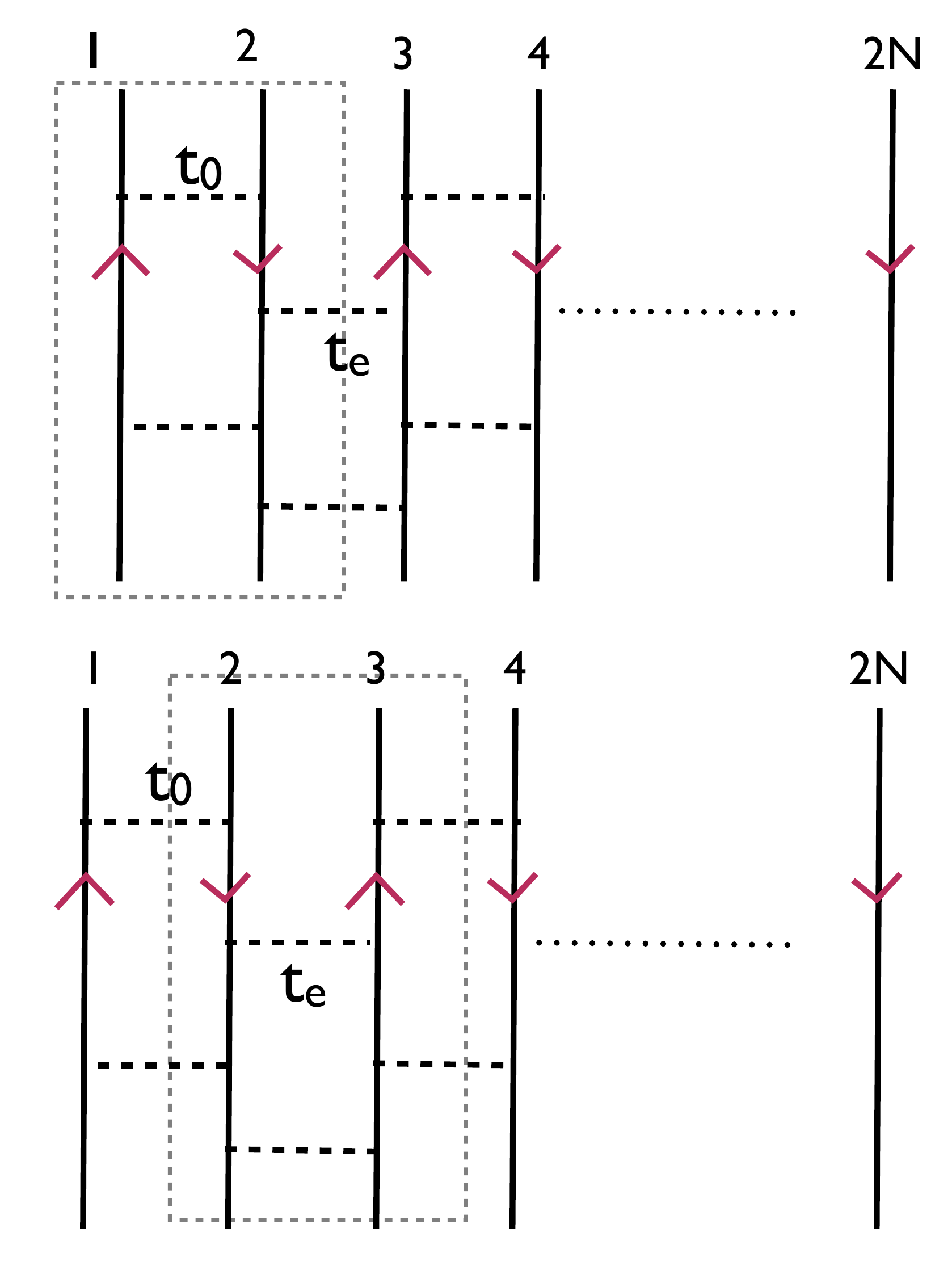}
\par\end{centering}
\caption{Network model for the fermion IQHE transition. On the top all chiral edge modes are paired to yield an ordinary insulator. On the bottom there is an unpaired edge mode to yield an integer quantum Hall insulator }
\label{nwrkpic}
\end{figure}

We will study the phase transition using a ``network" model construction. The idea is to start with the theory of the edge state and couple together opposite edges. Let us warm up with the familiar example of the IQHE transition of fermions from a state with $\sigma_{xy} = 1$ to one with $\sigma_{xy} = 0$. The model is defined by Fig. \ref{nwrkpic}. It is described by the Euclidean action ${\cal S} = \int dx dt {\cal L}$ with
\begin{equation}
{\cal L} = \sum_i \bar{c}_i\left(\partial_\tau - i s_i \partial_x \right) c_i - \sum t_i \left(\bar{c}_{i+1} c_i + \bar{c}_i c_{i+1} \right)
\end{equation}
with
\begin{eqnarray}
t_{i even} & =  & t_e \\
t_{i odd} & = & t_o \\
s_i & = & -(-1)^i
\end{eqnarray}
The first term is the sum of the actions of a single chiral edge mode of the $\sigma_{xy} = 1$ fermion IQHE state taken to propagate in opposite directions for adjacent $i$.
The second term describes electron hopping between opposite moving edge channels.

When $t_e < t_o$ all chiral edge channels are paired with partners and $\sigma_{xy} = 0$. Conversely if $t_e > t_o$, then all edge channels get paired except the two end ones and we get the fermion IQHE state with $\sigma_{xy} = 1$. The transition occurs at $t_e = t_o$ and is readily seen to yield a single massless Dirac fermion in the continuum limit is taken in the $i$ direction.

Let us now repeat this construction for the bosonic IQHE transition. The edge theory for the boson IQHE state with $\sigma_{xy} = 2, \kappa_{xy} = 0$ has  a pair of counterpropagating edge modes - one carries the charge and the other is neutral\cite{luav2012,tsml,WenSU(2)}. It is convenient to write the effective action of the edge as an $SU(2)_1$ WZW theory:
\begin{equation}
S_{eff} = \int dxd\tau \frac{1}{2\lambda} tr\left(\partial_\mu g^\dagger \partial_\mu g \right) + i S_{WZW}[g]
\end{equation}
Here $g$ is a $2 \times 2$ matrix with entries $g = \begin{pmatrix} b_1 & -b_2^* \\ b_2 & b_1^* \end{pmatrix}$. The $b_1, b_2$ are the two physical boson fields which form the IQHE quantum Hall state.

\begin{figure}
\begin{centering}
\includegraphics[scale=0.3]{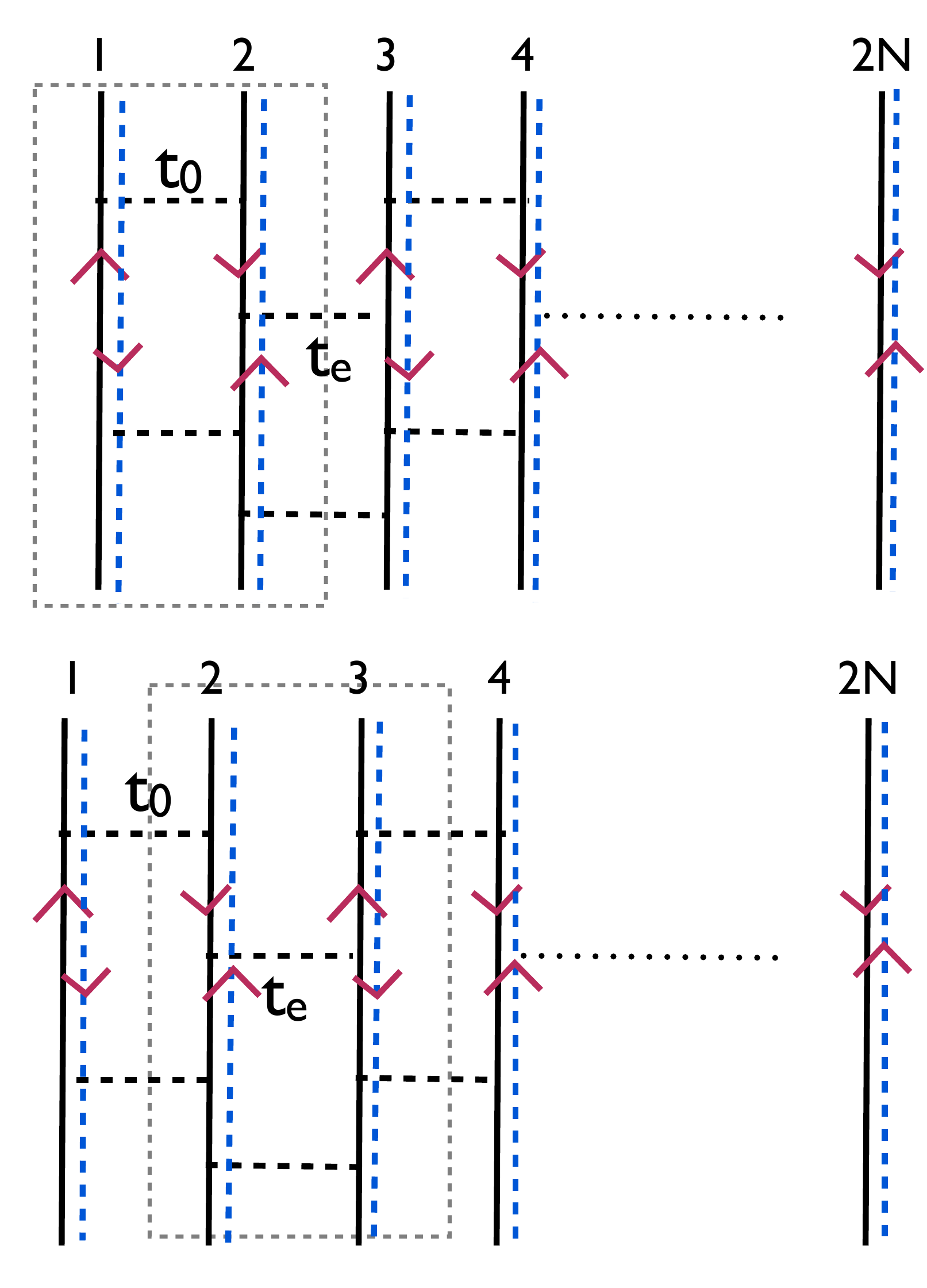}
\par\end{centering}
\caption{Network model for the boson IQHE transition. Each edge channel now has a charged chiral mode and a counter-propagating neutral mode. The rest is the same as for fermions.  }
\label{nwrkpicb}
\end{figure}

A network model capable of describing the boson IQHE state may now be written down and is defined by Fig \ref{nwrkpicb}.  Again we have an array of opposite edge channels which are coupled together by boson hopping $-\sum_{a = 1,2}\left(b^\dagger_{ia} b_{i+1,a} + h.c \right) \propto -tr\left(g^\dagger_{i}g_{i+1} + h.c\right)$. The full effective action is then
\begin{eqnarray}
S & = & S_0 + S_W + S_t \\
S_0 & = & \int dxd\tau \frac{1}{2\lambda}\sum_i tr\left(\partial_\mu g^\dagger_i \partial_\mu g_i \right) \\
S_W & = & i\sum_i s_i S_{WZW}[g_i] \\
S_t & = & - \sum_i t_i tr\left(g^\dagger_{i}g_{i+1} + h.c\right)
\end{eqnarray}
with $s_i$ and $t_i$ as before. If $t_o \gg t_e$ we get the trivial insulator while if $t_e \gg t_o$ we get the boson IQHE state. The transition occurs at $t_e = t_o$. A low energy theory of the transition is obtained by taking the continuum limit in the $i$ direction. As the opposite moving edge channels have opposite WZW terms they nearly cancel, and it is necessary to carefully sum them. Fortunately precisely this sum was performed in Ref. \onlinecite{tsmpaf2006} where the same model arose in a different context. The result is the effective $D = 2+1$ dimensional action
\begin{equation}
S_{eff} = \int d^3x \frac{1}{2\kappa}tr\left(\partial_\mu g^\dagger \partial_\mu g \right) + i \pi {\cal L}_\theta [g]
\end{equation}
The second term is a $\theta$ term for the $SU(2)$ matrix valued field $g$ in $2+1$ dimensions corresponding to $\Pi_3[SU(2)] = Z$. In the present context our calculation has yielded this term at the value $\theta = \pi$.

We of course do not have full $SU(2)$ symmetry rotating between $b_1$ and $b_2$ in the microscopic system. For the time being let us assume that we have $U(1) \times U(1)$ symmetry corresponding to separate conservation of the $b_1, b_2$ bosons. Further let us also assume there is a $Z_2$ symmetry interchanging $b_1$ and $b_2$. Later we will relax all these assumptions. Then the results of Ref. \onlinecite{tsmpaf2006} show that the field theory above at $\theta = \pi$ maps to the self-dual easy plane non-compact $CP^1$ (NCCP$^1$) model. Equivalently it also maps onto a model of two species of spacetime loops with a phase $\pi$ associated with each linking of the two loop species.

The $\theta = \pi$ $SU(2)$ matrix field theory (with the $U(1) \times U(1)$ anisotropy) or the equivalent easy plane NCCP$^1$ model arise also in the theory of deconfined quantum criticality in two space dimensions. Remarkably we see that the field theories describing the boson IQHE plateau transition  (and hence the surface states of the 3d boson topological insulator) are closely related to the theory of deconfined quantum criticality.  In the previous subsection we obtained this connection from a different point of view.  We will in the rest of the paper explore this connection in more detail and generality . For now we merely point out that the results of Ref. \onlinecite{tsmpaf2006} (see also Ref. \onlinecite{xul12}) show that the $\theta = \pi$ $SU(2)$ matrix field theory in two space dimensions does not have a trivial gapped disordered phase. Its phases either break symmetry, are gapless or have topological order. This is a hallmark of the surface state of a symmetry protected topological phase - there is no trivial gapped phase that preserves all the symmetries.

{\bf Surface of the Bosonic TIs: Field theories}
We now exploit our intuition about deconfined criticality to obtain the theory of possible surface states of the 3d boson TI starting with the $\theta = \pi$ $SU(2)$ matrix field theory.
We first describe a number of equivalent field theoretic descriptions of the surface state paying particular attention to the realization of the physical $U(1) \rtimes Z^T_2$ symmetry.
First we note that the $SU(2)$ matrix $g$ is related to the physical boson fields $b_{1,2}$ through
\begin{equation}
g = \begin{pmatrix} b_1 & -b_2^* \\ b_2 & b_1^* \end{pmatrix}
\end{equation}
Under the global $U(1)$ symmetry, both bosons transform with charge $1$, {\em i.e}
\begin{equation}
b_{1,2} \rightarrow b_{1,2} e^{i\varphi}
\end{equation}
We implement time reversal by simply requiring that
\begin{equation}
b_{1,2} \rightarrow b_{1,2}
\end{equation}
In terms of the phases of the bosons, defined through $b_{1,2} \sim e^{i\phi_{1,2}}$, and the conjugate bosons densities $n_{1,2}$ the symmetry transformations are the same as in Eqns. \ref{phiTransform},\ref{nTransform} so that we are indeed describing the same symmetry class in the two approaches.
We remind the reader that the total number $n_1 + n_2$ of the two boson species is conserved due to the global $U(1)$ symmetry but the relative number $n_1 - n_2$ is in general not. As promised before we will first analyse the theory in a limit where this relative number is also conserved (so that there is $U(1) \times U(1)$ symmetry) and then include interspecies tunneling terms to recover the generic case.

 \begin{table}[htdp]
\begin{tabular}{|c|c|c|c|c|c|}
\hline
Field & $q_1$ & $q_2$ & $n_{v1}$ & $n_{v2}$ & $Z_2^T$ \\ \hline
$\psi^\dagger_{1+}$ & $\frac{1}{2}$ & $0$ &  $0$ & $1$ & $\psi_{1-}$ \\ \hline
$\psi^\dagger_{11}$ & $ -\frac{1}{2}$ & $0$ & $0$ & $1$ & $\psi_{1+}$ \\ \hline
$\psi^\dagger_{2+}$ &  $0$ & $\frac{1}{2}$ &  $1$ & $0$ & $\psi_{2-}$ \\ \hline
$\psi^\dagger_{2-}$ &  $0$ & $-\frac{1}{2}$ & $1$ &  $0$ & $\psi_{2+}$ \\
\hline
\end{tabular}
\caption{Symmetry properties of the $\psi_{1\pm}, \psi_{2\pm}$ fields. $q_{1,2}$ are the charges under the two $U(1)$ symmetries associated with $b_{1,2}$ respectively. $n_{v1,2}$ are the vorticities in the phase of $b_{1,2}$. They can also be viewed as the gauge charge for the coupling to the corresponding $U(1)$ gauge fields. The last column gives the transformation under time reversal.}
\label{symmtbl}
\end{table}%

As argued in Ref. \onlinecite{tsmpaf2006},  the $\theta$ term of the $SU(2)$ matrix field theory with $U(1) \times U(1)$ anisotropy has a simple interpretation. It is the phase that is picked up when the vortex of the boson $b_1$ is taken around the vortex of the boson $b_2$. At $\theta = \pi$ the two vortices are mutual semions. We may thus readily write down a dual field theory in terms of the vortices $\Phi_{1v,2v}$ of the two bosons $b_{1,2}$ respectively. This has the structure
\begin{eqnarray}
\label{svrtx}
{\cal L} & = & {\cal L}_{1v} + {\cal L}_{2v}+ {\cal L}_\theta +{\cal L}_A\\
{\cal L}_{iv} & = & |\left(\partial_\mu - i (a_{i\mu} + \beta_{i\mu})\right)\Phi_{iv}|^2 + ....... \\
{\cal L}_\theta & = & \frac{i}{\pi} \beta_{1\mu} \epsilon_{\mu\nu\lambda}\partial_\nu \beta_{2\lambda} \\
{\cal L}_A & = & i A_{i\mu} j_{i\mu}
\end{eqnarray}
Here $\mu, \nu, \lambda,....$ represent space-time indices in $2+1$ dimensions. The $a_i, i = 1.2$ are the usual dual gauge fields of the vortex theory. The physical current
$j_{i\mu}$ of the bosons $b_{1,2}$ is given as usual by
\begin{equation}
j_{i\mu} = \frac{1}{2\pi} \epsilon_{\mu\nu\lambda} \partial_\nu a_{i\lambda}
\end{equation}
We have included external probe gauge fields $A_{i\mu}$ that couple to these currents.
The $\beta_{i\mu}$ are `statistical' gauge fields that serve to impose the mutual statistics of the two vortex species through the mutual Chern-Simons term in ${\cal L}_\theta$. We have also tuned away a chemical potential that couples to the total boson number so that the effective action is relativistic. We will shortly relax that assumption.

In passing we note that recently related models of two species of bosons with mutual $\pi$ statistics have been studied numerically through Monte Carlo simulations\cite{lesik12a}.
The relevance of these models to the surface of the 3d boson topological insulator (and the related 2d boson integer quantum Hall transition) should give further impetus for such studies.

\subsection{Synthesis of the Two Approaches}
We now provide a synthesis of the results of the two approaches taken in this section.  We will rely closely on the results of Ref. \onlinecite{tsmpaf2006} to provide
two alternate field theoretic representations of the theory described by Eqn. \ref{svrtx}. Rather than repeat the derivation from Ref. \onlinecite{tsmpaf2006}, we provide a physical description. The $\pi$ phase picked up by the vortex $\Phi_{1v}$ goes around the vortex $\Phi_{2v}$ suggests that $\Phi_{1v}$ carries $1/2$ charge under the global $U(1)$ symmetry associated with species $2$ and vice versa ({\em i.e}, $\Phi_{2v}$ caries $1/2$ charge under the global $U(1)$ of species 1). However the $\pi$ phase is obtained for both charge $1/2$ and charge $-1/2$. We thus should expect that the vortex of either species carries fractional charge $\pm 1/2$ of the global $U(1)$ quantum number of the other species. This expectation is formalized by the derivation\cite{tsmpaf2006}. First by doing a duality on one species (say 1) we explicitly map to an easy plane non-compact $CP^1$ model with action given by Eqn. \ref{vortexaction2}.

If instead we had performed a duality transformation on species 2, we would have obtained an equivalent action in terms of the fractionalized fields $\psi_{1\pm}$ related to
$b_1$ through
\begin{equation}
b^\dagger_1 = \psi_{1+}^\dagger\psi_{1-}
\end{equation}
This action takes the form
\begin{eqnarray}
{\cal L}_{2} & = &
\sum_s \left |\left(\partial_\mu - i \alpha_{1\mu} -i\frac{sA_1}{2}\right)\psi_{1s}\right|^2 + .... \nonumber  \\
& & + \frac{1}{2\kappa_1} \left( \epsilon_{\mu\nu\lambda}  \partial_\nu \alpha_{1\lambda} \right)^2 + \frac{i}{2\pi}A_{2\mu}\epsilon_{\mu\nu\lambda} \partial_\nu \alpha_{1\lambda}
\label{vortexaction1}
\end{eqnarray}
with the physical $U(1)$ current of the $b_2$ bosons given by $j_{2\mu} = \frac{1}{2\pi}\epsilon_{\mu\nu\lambda}\partial_\nu \alpha_{1\lambda}$. Note the obvious similarity of Eqn. \ref{vortexaction1} with Eqn. \ref{vortexaction2} after interchange of the $1$ and $2$ labels.  This is a reflection of the self-duality of the easy plane NCCP$^1$ model first pointed out in Ref. \onlinecite{lesikav2004}.This self-duality is obvious when both theories are obtained starting with the sigma model or the equivalent  dual vortex theory (Eqn. \ref{svrtx}).

\subsection{Analysis of the Surface Field Theory: Phase Diagram, Deconfined Criticality and Fermionic Vortices}
\label{SurfaceAnalysis}
Having obtained a field theoretic description of the surface states we now analyse the phase diagram.
The symmetry transformations summarized in Table. \ref{symmtbl} enables us to deduce the allowed perturbations to the actions above. A crucial allowed perturbation are `chemical potential' terms that couple to the boson number $\mu_1n_1 + \mu_2 n_2$. Another crucial allowed perturbation is an interspecies boson tunneling term $-\lambda\left(b^\dagger_1 b_2 + h.c \right)$. Let us first discuss the phase diagram when these terms are tuned to zero. Depending on the question being asked we will find it useful to use one or the other of the formulations provided above. For clarity of presentation we will however use the theory in Eqn. \ref{vortexaction2} to the extent that is convenient.

{\bf  I. $\mathcal T$ Breaking States and Quantum Hall Effect}
Consider condensing just one of the vortex fields in Eqn. \ref{vortexaction2}.
\begin{equation}
\langle \psi_{2+} \rangle \neq 0;  \langle \psi_{2-} \rangle = 0
\end{equation}
Such a phase clearly breaks T-reversal symmetry, which interchanges the vortices. However, it is an insulator  because the gauge invariant combination $\langle \psi_{2+}^*\psi_{2-}\rangle =\langle e^{i\phi_2}\rangle =0$. The transport properties of this phase are readily obtained by noticing that the combination $\alpha_1 + \frac{A_2}{2}$ is Higgsed. Therefore at long wavelengths we may set
\begin{equation}
\alpha_2 \approx - \frac{A_2}{2}
\end{equation}
Further we may integrate out the field $\psi_{2-}$ in Lagrangian \ref{vortexaction2}. The effective long wavelength Lagrangian for the external probe gauge fields then becomes
\begin{equation}
{\cal L}_{eff} = -\frac{i}{4\pi} A_{1\mu}\epsilon_{\mu\nu\lambda} \partial_\nu A_{2\lambda}
\end{equation}
Defining the `charge' and `pseudospin' probe gauge fields $A_c = \frac{A_1 + A_2}{2}$, $A_s = \frac{A_1 - A_2}{2}$, we get
\begin{equation}
{\cal L}_{eff} = -\frac{i}{4\pi}\left(A_{c\mu}\epsilon_{\mu\nu\lambda} \partial_\nu A_{c\lambda}  - A_{s\mu}\epsilon_{\mu\nu\lambda} \partial_\nu A_{s\lambda}\right)
\end{equation}
This implies that the charge Hall conductivity $\sigma_{xy} = -1$ while the pseudospin Hall conductivity $\sigma^s_{xy} = 1$. Taken together the thermal Hall conductivity $\kappa_{xy} = 0$. If on the other hand we had condensed $\psi_{2-}$ without condensing $\psi_{1+}$ we would have found the time reversed partner with $\sigma_{xy} = 1$ and $\sigma^s_{xy} = -1$.

Consider now adding symmetry allowed perturbations to the action. The surface state described above is gapped and hence is unaffected by the chemical potential terms if they are weak. The interspecies tunneling term destroys conservation of pseuodspin  (= $n_1 - n_2$) and hence $\sigma^s_{xy}$ is no longer quantized. However the electrical and thermal Hall conductivities continue to be well defined and will have quantized values $\sigma_{xy} = \pm 1, \kappa_{xy} = 0$. This is exactly what we expected based on the general considerations of Section \ref{Sec:gentrnsprt} above.

Pictorially, these edge states may be understood by considering the bulk system on a solid sphere and   assuming that $\psi_{1+}$ is condensed on the top hemispherical surface while $\psi_{1-}$ is condensed on the bottom hemispherical surface. Then along the equator there is a domain wall between the two kinds of surface quantum Hall states. At this domain wall there will be gapless one dimensional states identical to the edge of the $2d$ boson IQHE state.  Specifically there is one charged chiral mode corresponding to the jump $\Delta \sigma_{xy} = 2$ across the domain wall and a counterpropagating neutral mode which carries the pseudospin. When interspecies tunneling is added the quantization of the pseudospin Hall conductivity is not guaranteed but the neutral edge mode is protected so long as charge is still conserved.

{\bf II. Superfluid State} Let us now consider $T$-reversal symmetric phases. A simple option is
\begin{equation}
\langle \psi_{2s} \rangle = \psi_0
\end{equation}
independent of $s$. This state has $<b_2> \neq 0$. The gauge field $\alpha_{2}$ is Higgsed by the $\psi_{2\pm}$ condensate.    Consider for a moment the situation where the boson number is independently conserved for each species. Then this state breaks the global $U(1)$ symmetry associated with $b_2$ but preserves the other global $U(1)$ associated with $b_1$.  We will refer to it as SF$_1$. If on the other hand both $\psi_{2\pm}$ are gapped then they may be integrated out to leave behind a Maxwell action for $\alpha_1$. Integrating out $\alpha_1$ then gives a Higgs mass for $A_1$ so that the global symmetry associated with $b_1$ is now broken. We will call this SF$_2$. These two phases are separated by a phase transition that is described by the putative critical point of the easy plane NCCP$^1$ field theory.
In general, a chemical potential term can also be added which will tune the system away from the NCCP$^1$ critical point. Apart from SF$_1$ and SF$_2$, we have the possibility of a phase with coexistence of the two superfluid orders.

Inclusion of interspecies tunneling has a more dramatic effect. First there is now no real distinction between SF$_1$ and SF$_2$ phases so that the phase boundary between them disappear. More importantly as the relative phase of $b^\dagger_1 b_2$ can no longer wind, vortices in $b_1$ are bound to vortices in $b_2$.  Note that $\psi_{1\pm}$ are vortices in $b_2$ and $\psi_{2\pm}$ are vortices in $b_1$. When we bind vortices in $b_1$ to vortices in $b_2$, the resulting vortices are created by fields
\begin{equation}
V^\dagger_{ss'} = \psi^\dagger_{1s} \psi^\dagger_{2s'}
\end{equation}
with $s, s' = \pm$. Note that for $s = - s'$ $V_{ss'}$ carries charge $0$ under the single remaining global $U(1)$ while for $s = s'$ it carries charge $\pm 1$. Thus the vortices no longer carry fractional charge. $V_{++}, V_{--}$ can be obtained as a composite of the boson creation operator and the  vortex $V_{+-}, V_{-+}$ so that only the latter are `elementary'. Furthermore $V_{+-}$ can mix with $V_{-+}$ due to the interspecies tunneling term. It follows that there is a unique elementary vortex $V \sim V_{+-}$ which carries charge $0$. Further under time reversal
\begin{equation}
V \rightarrow V^\dagger
\end{equation}
Thus in the presence of interspecies mixing there is a {\em unique} vortex which does not transform projectively under the global symmetries. Does this invalidate our earlier analysis? In particular can we now get a trivial insulator by condensing this vortex? The answer is no. The point which we demonstrate below is that the effective action for the vortex $V$ in the superfluid phase is not the usual one but rather contains an extra Chern-Simons term. The presence of this Chern-Simons term has a convenient rough interpretation. It changes the statistics of the vortex to a fermion! A simple way to picture this is in terms of
 the description in terms of the vortex fields $\Phi_{v1,2}$ in Eqn. \ref{svrtx}. The $\theta$ term in the sigma model description means that the two vortices are mutual semions. It follows that their bound state is a fermion.

To put some meat into this picture we start with Eqn. \ref{svrtx}. In the presence of interspecies tunneling the individual vortex fields $\Phi_{1v}, \Phi_{2v}$ will be confined but a bound combination
$\Phi_{cv} = \Phi_{1v} \Phi_{2v}$ will survive. It is therefore necessary  to reformulate the action in terms of $\Phi_{cv}$. It is convenient to do so first even in the presence of the enlarged $U(1) \times U(1)$ symmetry and later include the interspecies tunneling. To do this we introduce another field $\Phi_{sv} = \Phi_{1v} \Phi^*_{2v}$. The resulting Lagrangian takes the form
\begin{eqnarray}
{\cal L} & = & {\cal L}[\Phi_{cv}, a_+ + \beta_+] + {\cal L}[\Phi_{sv}, a_- + \beta_-] + \nonumber \\
& + & \frac{i}{4\pi}\beta_{+\mu}\epsilon_{\mu\nu\lambda}\partial_\nu \beta_{+\lambda} - \frac{i}{4\pi}\beta_{-\mu}\epsilon_{\mu\nu\lambda}\partial_\nu \beta_{-\lambda}
\end{eqnarray}
Here $a_{\pm} = a_1 \pm a_2$, $\beta_\pm = \beta_1 \pm \beta_2$. $a_{\pm}$ are the usual dual gauge fields whose curl gives the charge and pseudospin current respectively.
The most interesting terms are the coupling to the gauge fields $\beta_\pm$ which have self Chern-Simons interactions. Including interspecies tunneling leads to linear confinement of $\Phi_{sv}$. The effective dual Landau-Ginzburg  theory of the superfluid then has the usual form but with the additional Chern-Simons term as promised.

{\bf III. Surface Topological Order} A wide variety of other phases are possible depending on the details of the surface interactions. For instance a gapped topologically ordered $Z_2$ liquid is possible and is accessed within the present formulation by condensing the paired vortex $(\psi_{2+}\psi_{2-}+{\rm h.c.})$ without condensing any other fields. In this situation the full three dimensional system when placed on a solid torus will have a ground state degeneracy of $4$ coming from the surface topological order.
It is interesting to consider the properties of this state a bit more and its relationship with the superfluid state. In the $Z_2$ topologically ordered insulator there is an  unpaired vortex $\psi_{2+} \sim \psi_{2-}^*$ which survives as a gapped excitation and which carries physical boson charge $1/2$ of the $U(1)_2$ global symmetry associated with the boson $b_2$. We will refer to it as a 2-chargon. Following standard reasoning this phase may equivalently also be understood as a paired condensate of $\psi_{1+}\psi_{1-}+{\rm h.c.}$. Thus there  is another gapped excitation corresponding to the field $\psi_{1+} \sim \psi_{1-}^*$ which, in the present context also carries charge $1/2$ of the physical boson
$b_1$. We will refer to this as the 1-chargon. These two chargons are mutual semions as expected for $Z_2$ topological order. Note that they have bosonic self-statistics.In the presence of inter-species tunneling, a pair of 1-chargons can mix with a pair of 2-chargons. Both species of chargons continue to exist as independent excitations but now they carry charge-$1/2$ of the remaining global $U(1)$. Finally the bound state of these two kinds of chargons is a fermion which does not carry fractional charge. We can take it to be charge neutral. It is convenient to regard this neutral fermion as the vison, and the two kinds of bosonic chargons as the other two non-trivial quasiparticles expected for a $Z_2$ topological ordered state. These transformation laws are summarized in Table\ref{table:Z2symm}.

Now let us consider the relationship to the superfluid state discussed above. This will enable us to clarify the nature of the vortices of the superfluid state. Coming from the superfluid side the $Z_2$ topological state is obtained by condensing paired vortices. In the presence of inter-species tunneling we argued above that there is a unique vortex $V$. The unpaired vortex survives as a finite energy vison in this vortex pair condensate. That this vison is a fermion ties in nicely with the observation that the superfluid vortex $V$ is conveniently regarded as a fermion. Thus as the transition to this $Z_2$ insulator is approached the vortex statistics becomes well defined and becomes fermionic.

The topologically ordered phase provides a particularly simple perspective on why a trivial gapped paramagnet is not allowed. Generally to go from a topologically ordered insulator to a trivial insulator we must confine the topological quasiparticles. For a $Z_2$ gauge theory, this is done b condensing one of the three non-trivial kinds of quasiparticles (usually dubbed the electric, magnetic and their composite). For the $Z_2$ topological state that can appear at the surface of the SPT phase we are discussing, the electric and magnetic particles are both (half)-charged under the global $U(1)$ symmetry, and their condensation breaks this symmetry. On the other hand the neutral topological quasiparticle is a fermion and hence it cannot condense. At the same time, time reversal prevents one from altering the Chern number associated with this gapped fermion. Thus we see clearly that a trivial gapped state obtained by confinement from the $Z_2$ topological state is not possible at the surface.

An interesting property of this $Z_2$ topological ordered state is that it realizes symmetry differently from strictly $2d$ systems. Such a strictly two dimensional gapped abelian insulator may be described within the usual $K$-matrix formulation. For $Z_2$ topological order $K = 2\sigma^x$. If both bosonic chargons carry charge $1/2$ as we argued then the charge vector $\tau = (1, 1)$. It is then easy to see that the resulting topological phase has non-zero electrical  Hall conductivity, and therefore must break time reversal invariance. However when realized at the surface of the three dimensional insulator a time reversal symmetric  $Z_2$ topological phase where both bosonic quasiparticles carry charge-$1/2$ is allowed.  In Appendix \ref{App:Z2}, we collect together the properties of the $Z_2$ topological ordered state for the various SPT phases discussed here, and show using the results of Ref. \onlinecite{LevinStern} that they all realize symmetry differently from what is allowed in strictly $2d$ systems.

While such interesting topologically ordered (or other even more exotic) states are allowed they are not required:  the surface could be in a superfluid or $T$-broken insulating quantum Hall state with no ground state degeneracy. The most important conclusion however is that a trivial gapped insulating state which preserves all the symmetries and has no topological order is not possible on the surface. This is a key property of a symmetry protected topological phase and is satisfied by our example.

 \begin{table}[htdp]
\begin{tabular}{|c|c|c|}
\hline
Excitation & Charge & ${\mathcal T}^2$ \\ \hline
Boson 1 (e) & q=1/2\, & +1 \\ \hline
Boson 2 (m) & q=1/2 \, & +1 \\ \hline
Fermion (f) &  q=0 \,\, & +1 \\
\hline
\end{tabular}
\caption{Symmetry properties of the topological excitations of a $Z_2$ gauge theory, realized on the surface of a 3D SPT phase with charge conservation and time reversal symmetry ($U(1)\rtimes Z_2^T$), as for the topological insulator. The bosonic quasiparticles carry half charge, but transform linearly under time reversal. Further details are in Appendix \ref{App:Z2}
\label{table:Z2symm}}
\end{table}%

Some remarks are now in order. We have constructed one topological phase, which when the surface spontaneously breaks time reversal symmetry leads to a quantized magneto electric effect.  The cohomology classification of Chen et al.\cite{chencoho2011} gives $Z_2^2$ so that there are three non-trivial states. Of the these three, one must have vanishing magneto-electric effect since this is an additive quantity.  We expect that this phase is simply the SPT phase associated with just the $Z_2^T$ symmetry discussed in Section \ref{Sec:TimeRev}. The remaining phase is then obtaned by combining the other two phases.   In the particular example discussed in this Section, the fact that there is a single U(1) charge conservation symmetry introduces additional terms that imply a deformation of the deconfined criticality action.  Despite this the degrees of freedom of this action provide the useful fields in terms of which the effective theory of the surface state may be described. We will discuss other examples with different symmetries in subsequent sections.


\section{3D Topological Field Theories}
\label{Sec:3D}
We would like to write down a 3D theory of particles and their vortices. We can choose to represent particle currents by $j^\mu =\epsilon^{\mu\nu\lambda\sigma}\partial_\nu B_{\lambda\sigma}/2\pi$. The vortex lines, being loops in three dimensional space, sweep out a surface in space time defined by the two form $\gamma^{\mu\nu}$. Relating this to a vector potential $a$, whose curl is the location of the vortex loop, we define $\gamma^{\mu\nu} = \epsilon^{\mu\nu\lambda\sigma}\partial_\lambda a_{\sigma}/2\pi$. The quantization of boson particle number to integers implies that the charges that couple to this vector potential $a$ are quantized. (Equivalently, the dual vector potential is compact, i.e. only defined modulo $2\pi$). Clearly, gauge transformations $a_\mu \rightarrow a_\mu+\partial_\mu \chi$ and $B_{\mu\nu} \rightarrow B_{\mu\nu}+(\partial_\mu \alpha_\nu -\partial_\nu\alpha_\mu)$ do not change physical variables.   Taking a particle around a vortex leads to a phase of $2\pi$, which is captured by the minimal coupling ${\mathcal L}= a_\mu j^\mu$, which may be rewritten as:

\begin{equation}
{\mathcal L}= \frac1{2\pi} a_\mu\epsilon^{\mu\nu\lambda\sigma}\partial_\nu B_{\lambda\sigma}
\end{equation}

This is often written as $\epsilon B\partial a/2\pi$, and called the {\em BF} action. The unit coefficient in the action ensures that there is no topological order, i.e. a unique ground state in the absence of surfaces \footnote{With an integer coefficient $Q$, ${\mathcal L}^E_b= \frac{Q}{2\pi}\epsilon B \partial a$, describes a topologically ordered phase with  ground state degeneracy of $Q^3$ on a three torus. } - appropriate to the current discussion. Then, the theory above only states the obvious, that particles and their vortices have a mutual phase factor of $2\pi$.

Let us briefly describe how this term arises from a microscopic theory of bosons. A lattice regularized theory of bosons can be captured by a loop model of integer valued closed loops with Euclidean Lagrangian ${\mathcal L}_b = \frac1{2\rho} {j_\mu j_\mu}$. The  integer constraint is implemented by summing over the auxiliary vector field $a_\mu$ that is an integer multiple of 2$\pi$. Now, the current $j_\mu$ takes real (rather than integer) values, and its divergence free condition can be implemented by writing $j_\mu= \epsilon_{\mu\nu\lambda\sigma}\partial_\nu B_{\lambda\sigma}/2\pi$, where the two form $B$ is also a real field. This gives:
\begin{equation}
{\mathcal L}^E_b = \frac1{8\pi^2\rho} (\epsilon \partial B)^2 + \frac{i}{2\pi} \epsilon a \partial B
\label{Lb}
\end{equation}
where the second term is the desired statistical interaction (the factor of $i$ appears because of the Euclidean formulation). However, at this point $a$ is an integer (times $2\pi$) field. One can softly introduce this constraint by assuming $a$ to be real, but adding the cosine term $\Delta {\mathcal L} = -\lambda \cos (\partial_\mu \phi - a_\mu)$, where we utilized the fact that longitudinal component can always be added to the gauge field \cite{FisherLee}. The phase $\phi$ is actually the phase of the original bosons, and when the bulk is insulating the cosine is irrelevant, since the bosons are gapped. Therefore in the insulating phase we may use Eqn. \ref{Lb} where both $B$, $a$ fields taken real, with the caveat that charges are ultimately quantized. Further discussion of basic issues related to this theory is at the end of Appendix A.

As shown in appendix A, surface states defined from the BF theory Eqn. \ref{Lb} are usual 2D bosonic modes that are not topologically protected.  To encode a SRE topological phase an additional term must be added as shown below.

Based on the discussion on surface states, where a pair of bosonic fields were invoked, we consider two species of bosons to write down a topological term. This also follows the two component U(1)$\times$U(1) symmetric Chern Simons approach for the 2D systems, which was found to be successful in describing 2D SRE topological phases\cite{luav2012}. Therefore we will introduce two $B$ fields that represent their conserved currents, and two $a$ fields which are vortices in these fields. Let us begin with the general case of $N$ species of bosons, with:

\begin{equation}
{\mathcal L}_{BF} = \frac{1}{2\pi}\sum_{I=1}^N  \epsilon B_I\partial a_I
\label{BdA}
\end{equation}
where $\epsilon$ is the antisymmetric symbol and indices have been suppressed.

Note, the apparently more general version is ${\mathcal L}_B = \frac{Q_{IJ}}{2\pi}\epsilon B_I\partial a_J$ with $Q$ a uni-modular i.e. $\det Q=1$ integer matrix which ensures absence of topological order. However, this can be brought into the canonical form of Eqn\ref{BdA} by redefining $B_I= [Q^{-1}]_{KI}B'_K$. The transformation matrix $Q^{-1}$ is also an integer matrix, since $\det Q=1$ and the minors of an integer matrix are also integers.

Now, an additional topological term can be added to the action:
\begin{eqnarray}
{\mathcal L}_{3D} &=&{\mathcal L}_{BF} +{\mathcal L}_{FF} \nonumber\\
{\mathcal L}_{FF} &=&  \frac{\Theta}{8\pi^2}K_{IJ}\epsilon\partial a_I\partial a_J
\label{Eqn:FF}
\end{eqnarray}
The action must be invariant under $\Theta\rightarrow \Theta+2\pi$, to allow for addition of 2D layers at the surface. Therefore, the action defined on a closed three dimensional space should be invariant  under the shift $\Theta\rightarrow \Theta+2\pi$. It is shown in Appendix \ref{Kinteger} that this condition fixes the entries $K_{IJ}$ to be integers.

A stronger condition on $K$ can be applied as follows. Values of $\Theta$ that differ by $2\pi$ simply correspond to different ways of terminating the surface. Hence, at a domain wall where $\Theta$ changes by $2\pi$ at the surface, we demand that all excitations present are bosonic. This is the same as the requirement placed on $K$ matrices describing 2D SPT phases, i.e. that  ${\rm Det} K=1$ and all diagonal entries are even integers.

The simplest choice of $K$ matrix with these properties is:
\begin{equation}
K=
\left(
\begin{array}{cc}
 0   & 1   \\
  1 & 0
\end{array}
\right)
\label{K}
\end{equation}
We find that for most of the 3D bosonic SPT phases that we will be interested in, it will suffice to consider this matrix. This is similar to the 2D situation\cite{luav2012}, where the above $K$ matrix describes a large set of SPT phases, which differ in the way symmetry is implemented.

\subsection{Two component BF theory of Bosonic Topological Insulator}Let us specialize to the symmetries of the topological insulator $U(1) \rtimes Z_2^T$, and consider a two component theory with the simplest allowed $K$ matrix given by Eqn. \ref{K}. Then we can write:
\begin{eqnarray}
{\mathcal L}_{tot}&=&{\mathcal L}_{BF}+{\mathcal L}_{FF}+{\mathcal L}_{em}\\
{\mathcal L}_{BF} &=& \frac{1}{2\pi} \epsilon \left ( B_1\partial a_1+B_2\partial a_2 \right ) \\
{\mathcal L}_{FF} &=&  \frac{\Theta}{4\pi^2}\epsilon\partial a_1\partial a_2\\
\end{eqnarray}
We will discuss coupling to the external electromagnetic field ${\mathcal L}_{em}$ subsequently.  Under time reversal symmetry we have:
\begin{eqnarray}
B_{I,0i}\rightarrow -B_{I,0i};\,&&B_{I,ij}\rightarrow B_{Iij}\\
a_{I,0}\rightarrow a_{I,0};\,&&a_{I,j}\rightarrow -a_{I,j}
\end{eqnarray}
where indices $i,j$ refer to spatial coordinates. The transformation of $B$ fields is obtained by relating them to the boson densities and currents, while the $a$ fields are chosen to transform such that the $BF$ term is left invariant. Since both species $a_I$ transform in the same way under time reversal we may conclude, $\Theta\rightarrow -\Theta$ under $Z_2^T$. A time reversal invariant bulk action can then be constructed for $\Theta=0,\,\pi$, (given the ambiguity in $\Theta$ modulo $2\pi$). We of course pick  $\Theta=\pi$ in the topological phase. While we have not derived this action, we have written down the simplest possible topological theory which meets the general constraints required of SPT phases. We now proceed to show it produces a surface with the same physical properties as predicted in the previous section. We study three different situations: first, we study the surface superfluid, and determine the quantum numbers of vortices. Second, we investigate the electromagnetic response, particularly the magneto-electric polarizability. Finally, we analyze the case where time reversal is broken at the surface, in opposite ways, leading to a domain wall.

{\em (i) Fractionally Charged Vortices:} Consider a surface of the topological phase at $z=0$ with $\Theta=\pi \,(0)$ for $z>0\, (z<0)$. Then the effective action at the surface arising from ${\mathcal L}_{FF}$ is:

\begin{equation}
S_{edge} = \frac1{4\pi}\int dtdxdy \, \epsilon^{z\alpha\beta\gamma} a_{1\alpha}\partial_\beta a_{2\gamma}
\label{edgevortex}
\end{equation}
where indices $\alpha,\beta,\gamma$ run over $t,x,y$, and the fields are evaluated at $z=0$.  \footnote{This is obtained by applying Gauss law $\int dV\partial_\mu j^\mu=\oint dS_\mu j^\mu$, where $j^\mu=\epsilon^{\mu\nu\lambda\sigma}a_{1\nu} \partial_\lambda a_{2\sigma}$}.
At the surface we may replace $a_{Ii}=\partial_{i} \phi_I$ (see appendix \ref{appendixBFsurface}), where $i=x,\,y$.  Consider now a surface superfluid of component $I=2$, with a vortex at the origin $x=y=0$. This implies $\phi_2$ winds around the origin, or that $(\partial_x\partial_y-\partial_y\partial_x )\phi_2=2\pi n^v_2\delta(x)\delta(y) $, where we have allowed for a vortex of strength $n^v_2$. Substituting this in Eqn. \ref{edgevortex} we have:

\begin{equation}
{\mathcal S}_a= \frac{1}{2}\int dt n^v_2 {\partial_t \phi_1}
\label{vortexmode}
\end{equation}

similarly, a vortex in the field $\phi_1$ couples to the phase of $\phi_2$.  Given the conjugate relation between number and phase, this implies that a vortex of strength $n^v_2$ in component $I=2$ carries charge $n^v_2/2$ of component $I=1$. Thus we see that unit vortices in one bosonic field carry a half charge of the other field.

A different perspective on this result is obtained by thinking about the fate of `external' monopoles of the gauge fields $a_1, a_2$. These are sources for vortex lines of the two boson fields $b_1$ and $b_2$ respectively. Now the well known Witten effect implied by the $\Theta$ term tells us that a $2\pi$ monopole in $a_1$ carries gauge charge
$\frac{\Theta}{2\pi} = \frac{1}{2}$ that couples to $a_2$, {\em i.e} it carries charge $1/2$ of the global $U(1)$ associated with $b_2$. Similarly a $2\pi$ monopole in $a_2$ carries a charge $1/2$ of the global $U(1)$ associated with $b_1$. At the surface this monopole creates a $2\pi$ vortex of the corresponding boson which is then seen to carry half-charge of the other boson thereby recovering the result above.

{\em (ii) Electromagnetic Response:}
Since one has U(1) symmetry, we can couple to an external electromagnetic field and write the following terms:

\begin{eqnarray}
{\mathcal L}_{em} =\frac{\epsilon}{2\pi}(q_1B_1\partial A_1+q_2B_2\partial A_2)
\label{em1}
\end{eqnarray}

Where $q_I$ is the charge on the $I$th boson given that the current of bosons $\epsilon \partial B_I =j$ couples minimally to $A_I$. If a single $U(1)$ charge is present, then we will identify $A_1=A_2=A$. We will see that this minimal coupling, along with Eqn. \ref{Eqn:FF}ensures the charged vortex ends are also coupled to the external electromagnetic field.

To find the electromagnetic response, we integrate out the $B$s, and then the $a$. The first step gives $a_I=-q_I A_I$. Substituting this we get:


\begin{eqnarray}
{\mathcal L}_{em-response}&=& \frac{\Theta}{4\pi^2}[q_1q_2]\epsilon\partial A_1\partial A_2
\label{3dEM}
\end{eqnarray}

Setting $\Theta=\pi$,$q_1=q_2=1$ this is:
\begin{equation}
{\mathcal L}_{em-response}= \frac{1}{4\pi}\epsilon\partial A_1\partial A_2
\end{equation}
This indicated that ends of vortices in the condensate of the first species of bosons (which are induced by inserting $2\pi$ flux in $A_1$) carry a half charge of the second species.

Finally, to obtain the response to the external electromagnetic field, we identify $A_1=A_2=A$, and then  ${\mathcal L}_{em}= \frac{2\pi}{4\pi^2}\vec{E}\cdot \vec{B}$. This corresponds to a magneto-electric polarizability $\theta=2\pi  $, i.e. it is an odd multiple of $2\pi$,  as expected.

The theory above can also be derived using the hydrodynamic approach of \cite{RyuFradkin}. Note, we could absorb the `FF' term in the `BF' term\cite{RyuFradkin} by redefining $B_I\rightarrow B_I+\frac{\Theta}{4\pi^2} K_{IJ}\epsilon \partial a_J$. However, this leads to an additional electromagnetic coupling: $\frac{\Theta}{4\pi^2}q_IK_{IJ}\epsilon\partial A_I\partial a_J$. However, we will prefer to work with the original form since it is a field theory written in terns of internal fields, that is well defined even in the absence of a conserved charge and external electromagnetic couplings.

{\em Surface Domain Wall:}
Finally, we consider an insulating surface in the $z=0$ plane, on which time reversal is broken in opposite ways for $y>0,\, y<0$. This leads to edge modes along the $x$ direction, localized near $z=y=0$. This is modeled with a spatially varying $\Theta$ field, where $\Theta(z>0)=0$, while $\Theta(z<0,y)=\pi {\rm sign} (y)$. Introducing this profile in ${\mathcal L}_{tot}$, with gapped $B$ field on the surface, the edge theory is readily shown to be:

\begin{eqnarray}
{\mathcal S}_{domain-wall}&=&\frac1{2\pi}\int dtdx\,  [ \partial_x\phi_1\partial_t\phi_2 \\ &&+A_0 \partial_x(q_2\phi_1+q_1\phi_2)+\dots ]
\end{eqnarray}
where the first term defines the commutation relations of a regular Luttinger liquid, the second term identifies the coupling to the external field (assuming a gauge where $A_x=0$), and the dots refer to non universal potential terms for the edge fields. This is identical to the edge state of the Integer quantum Hall effect of bosons \cite{luav2012,tsml}, on setting $q_1=q_2=1$, which is also constant with the magneto electric polarizability of $\theta=\pi$ in this phase.

The general problem of deriving the 3D field theory above, from microscopic models is left for future work. Below, we describe a bulk non-linear sigma model which is some ways may be considered a microscopic theory since it assume additional ingredients, over and above the purely symmetry group based approach of the cohomology theory\cite{chencoho2011}. Therefore, instead of writing  sigma models with topological terms, where the target manifold is the symmetry group, we will allow the target manifold to be the four sphere $S^4$, which assumes a particular microscopic representation. However, since we are not concerned with classifying phases, but rather with providing physical examples, this additional assumptions will be convenient. This is quite analogous to the common practice of considering crystalline band structures for free fermion topological insulators, although they can (and strictly speaking should) be defined in the absence of translation symmetry\cite{KitaevRyu}.

 \subsection{Bulk sigma model theory}
 The bulk field theory discussed in the previous section is topological and has no bulk dynamical degrees of freedom. In this section we describe a different bulk theory with dynamical boson fields which gives rise to the topological effective field theory of the previous section in a disordered phase. This theory may thus be viewed as a field theory realization of a model with a bosonic SPT phase in three space dimensions. This field theory takes the form of a $3+1$ dimensional non-linear sigma model supplemented with a topological $\theta$ term. This generalizes to three space dimensions the  continuum field theory model that realizes the $2d$ integer quantum Hall state of bosons.

Following the $2d$ example and the discussion in previous sections of this paper we will enlarge the symmetry of the boson system from $U(1)$ to a larger symmetry group and then add perturbations to reduce to the symmetry of interest. For the construction of this section it is extremely convenient to consider a generalization where we first embed the $U(1)$ symmetry into an $SO(5)$ group. Consider therefore a $5$-component unit vector field $\hat{n}$. Later we will describe exactly how the physical symmetry ($U(1) \times Z^T_2$ or $U(1) \rtimes Z^T_2$, etc) are realized by the components of this field. For now we write down a continuum field theory for $\hat{n}$. On a closed space-time manifold, say the four sphere, the Lagrangian takes the form
\begin{eqnarray}
{\cal L} & = & {\cal L}_0[\hat n] + {\cal L}_{\theta}[\hat n] \\
{\cal L}_0[\hat n] & = & \frac{1}{2\lambda} \left(\partial_\mu \hat n \right)^2 + ....  \\
{\cal L}_\theta[\hat n] & = & i\theta Q \\
Q & = &  \frac{1}{\Omega_4}\int d^3x d\tau \epsilon_{abcde} n_a \partial_x n_b \partial_y n_c \partial_z n_d \partial_\tau n_e \\
& = & \frac{3}{8\pi^2} \int d^3x d\tau \det[ \hat{n}~\partial_x \hat{n}~\partial_y \hat{n}~\partial_z \hat{n}~ \partial_\tau \hat{n} ]
\end{eqnarray}
where $\Omega_4 = \frac{8\pi^2}{3}$ is the volume of the unit four dimensional sphere. $Q$ is the integer invariant corresponding to $\Pi_4(S^4) = Z$ and counts the number of times spacetime configurations of the $\hat{n}$ field wrap around unit 4-sphere. Clearly the theta term does not affect the bulk physics if $\theta = 2n\pi$ with integer $n$. We note that through out this section $\theta$ will denote the
theta angle for the bulk sigma model, and should not be confused with the same symbol used previously for the electromagnetic response.
We are interested in disordered phases of this field theory where $\hat{n}$ is gapped in the bulk.

Consider a spatial domain wall between a state where $\theta = 2\pi$ and one where $\theta = 0$. If this domain wall has a non-trivial surface state then the $\theta = 2\pi$ theory describes an SPT phase in the bulk. Such a domain wall corresponds to a situation where $\theta$ varies spatially. To handle this it is convenient to elevate $\theta$ to be a new dynamical field and define a sigma model  for a new 6-component unit vector field $\hat{\phi}$ defined by
\begin{equation}
\hat{\phi} =
\left(
\begin{array}{cc}
&  \cos \alpha  \\
  & \hat{n}\sin \alpha
\end{array}
\right)
\end{equation}
The field $\phi$ defines a map from spacetime (taken to be the four dimensional sphere $S^4$) to the five dimensional unit sphere $S^5$. Consider a field theory for $\hat{\phi}$ which includes (apart from the usual gradient terms) a Wess-Zumin-Witten (WZW) term defined as usual as $2\pi$ times the fraction of the volume of $S^5$ that is bounded by the hypersurface traced out by $\hat{\phi}$. Formally let $\hat{\phi}(x,u)$ be a smooth extension of $\hat{\phi}(x)$ such that $\hat{\phi}(x, 0) = \hat{\phi}_0$, $\hat{\phi}(x, 1) = \hat{\phi}$.
Then
\begin{equation}
{\cal S}_{WZW}[\hat{\phi}] = \frac{2}{\pi^2}\int_{\vec x, \tau}\int_0^1 du \det[\hat{\phi}~\partial_x \hat{\phi}~\partial_y \hat{\phi}~\partial_z \hat{\phi}~\partial_\tau \hat{\phi}~\partial_u \hat{\phi}]
\end{equation}
Consider an extension where
\begin{equation}
\hat{\phi} =
\left(
\begin{array}{cc}
  &  \cos \alpha(u)   \\
  & \hat{n}\sin \alpha(u)
\end{array}
\right)
\end{equation}
with $\alpha(0) = 0$, $\alpha(1) = \alpha$, $\alpha(u)$ independent of $\vec x, \tau)$ and $\hat{n}$ is independent of $u$. The determinant  in ${\cal S}_{WZW}$ is readily calculated and reduces to the theta term ${\cal L}_\theta[\hat{n}]$ for the $\hat{n}$ field with
\begin{equation}
\theta = \frac{16}{3}\int_0^\alpha d\alpha' \sin^4 \alpha'
\end{equation}
In particular when $\alpha = 0$ we get $\theta = 0$ and when $\alpha = \pi$ we get $\theta = 2\pi$. Thus the WZW model for $\hat{\phi}$ with constant $\alpha = \pi$ describes the $3+1$-d non-linear sigma model for $\hat{n}$ at $\theta = 2\pi$.

To study a domain wall between $\theta = 2\pi$ and $\theta = 0$ along, say, $z = 0$ let $\alpha = \alpha(z)$ such that
\begin{eqnarray}
\alpha (z \rightarrow -\infty) & = & 0 \\
\alpha (z \rightarrow \infty) & = & \pi
\end{eqnarray}
Further we assume that $\frac{d\alpha}{dz}$ is localized to within a short distance of $z = 0$. To evaluate the WZW term in this configuration it is convenient to use a different extension of the $\hat{\phi}$ field. Specifically let
\begin{equation}
\hat{\phi} =
\left(
\begin{array}{cc}
  & \cos\alpha(z) \\
 &  \hat{n}(\vec x, \tau, u)\sin \alpha(z)

\end{array}
\right)
\end{equation}
with $\alpha$ now independent of $u, x,y, \tau$ and $\hat{n}(x, 0) = \hat{n}_0$, $\hat{n}_{x,1} = \hat{n}(x)$. The determinant in ${\cal S}_{WZW}$ is again readily evaluated and becomes
\begin{equation}
{\cal S}_{WZW} = \frac{2}{\pi^2} \int_{-\infty}^\infty dz \frac{d\alpha}{dz}\int_{x,u}\det[\hat{n}~\partial_x \hat{n}~\partial_y \hat{n}~\partial_z \hat{n}~ \partial_\tau \hat{n} ~\partial_u \hat{n}]
\end{equation}
As $\frac{d\alpha}{dz}$ is localized at the domain wall at $z = 0$, we can replace $\hat{n}$ in the integral by its configuration at $z = 0$. The $z$-integral can now be performed and leads to
\begin{equation}
{\cal L}_{WZW} = \frac{3}{4\pi}\int_{x,u}\det[\hat{n}~\partial_x \hat{n}~\partial_y \hat{n}~\partial_z \hat{n}~ \partial_\tau \hat{n} ~\partial_u \hat{n}]
\end{equation}
This is exactly the WZW term (at level $1$) for the $\hat{n}$ field at the boundary. Thus the domain wall in question is described by a $2+1$ dimensional  $SO(5)$ non-linear sigma model with a WZW term.

This field theoretic result is very useful to construct a bulk sigma model description of the  SPT phases discussed in this paper. The simplest application is to bosons with symmetry $U(1) \times Z^T_2$ discussed in detail in the next section. To illustrate this let us first introduce an $U(1) \times SO(3)$ anisotropy and write $\hat{n} = [ Re \psi, Im \psi, \vec N]$, where under the global $U(1)$ symmetry choose $\psi \rightarrow e^{i\epsilon} \psi$, but $\vec N  \rightarrow \vec N$. Under time reversal, we let $\psi \rightarrow \psi^* \vec N \rightarrow - \vec N$. Finally under the global $SO(3)$ symmetry $\vec N$ transforms as a vector while $\psi$ is invariant.

The level-1 WZW term plays the following crucial role\cite{tsmpaf2006} in this field theory: it implies that the vortex of the $\psi$ field transforms as spin-$1/2$ under the $SO(3)$ rotation.  Indeed  the 5-component sigma model with global $SO(3) \times U(1)$ and $Z^T_2$
symmetries implemented this way, and supplemented with a level-1 WZW term precisely arises also as the theory of the deconfined critical point between Neel and VBS states in $2d$.  There the spin-$1/2$ attached to the vortex captures the physical picture that a VBS vortex is a spinon.

For our present purposes we need to further explicitly break the $SO(3)$ symmetry while preserving time reversal. Then the $\vec N$ field is no longer a freely fluctuating variable.
However the crucial point is that as the vortices of $\psi$ form a spinor their Kramers degeneracy will be preserved so long as $Z^T_2$ is preserved even if the full $SO(3)$ is not present. This is exactly the defining property of the surface theory of one of the SPT phases for bosons with $U(1) \times Z^T_2$ symmetry described in the next section. We have thus obtained this surface theory from a bulk sigma model.

{\em (i) Meaning of theta term of bulk sigma model}

The sigma model description is useful as it suggests a route to obtaining a physical realization of this SPT phase. First
let us understand the meaning of the bulk $\theta$ term in this sigma model. As a topological term it depends on the global configuration of the $\hat{n}$ field.  In general for a theory of $\vec N, \psi$ with $SO(3) \times U(1)$ symmetry, if there are no topological defects in either the $\vec N$ or the $\psi$ field it is easy to see that the $\theta$ term vanishes. In $3d$ the $\vec N$ field admits point hedgehog defects while the $\psi$ field admits vortex loops. The $\theta$ term implies that during a process where a hedgehog is taken around a vortex line a phase $e^{i\theta}$ accumulates. For $\theta = 2\pi$ it follows that the hedgehog of $\vec N$ has charge $1$ under the global $U(1)$ symmetry associated with $\psi$ (this ensures that it acquires phase $2\pi$ when it moves around a vortex line). Thus this kind of SPT phase may potentially be engineered by constructing a physical situation where the hedgehog defect of a 3-component order parameter is charged under the global $U(1)$ symmetry of the bosons of interest.

{\em (ii) Other SPT phases}

To describe the other SPT phase discussed above with the same symmetries, we must implement the symmetry differently. First we consider a different situation where
the $SO(3)$ vector only has $U(1) \times Z_2^T$ symmetry; then we write $\hat{n} = (Re \psi_1, Im \psi_1, Re\psi_2, Im\psi_2, N_z)$. Following the previous sections we take
both $\psi_1$ and $\psi_2$ to be charged under the global $U(1)$ symmetry.  For $U(1) \times Z^T_2$ we take under time reversal $\psi_{1,2} \rightarrow \psi_{1,2}^*$ and $N_z \rightarrow - N_z$. If time reversal is preserved, then  $\langle N_z \rangle = 0$.  In the presence of anisotropy that favors the $\hat{n}$ to have zero component of $N_z$ we may drop it to obtain an effective field theory for the surface. Then the level-1 WZW term for $\hat{n}$ becomes the familiar $\theta$ term at $\theta = \pi$ for the remaining 4-components in the $2+1$ dimensional surface theory. To understand the bulk, note that when there is $U(1) \times Z_2^T$ anisotropy on an $SO(3)$ vector the vortex lines of the $U(1)$ field $\psi_2$ come in 2 kinds which are distinguished by the sign of $N_z$ in the core. The hedgehogs of the original $\vec N$ field are then domain walls within the cores of these vortices where the $N_z$ changes sign\cite{lesik05}.  Thus the hedgehog must be regarded as a composite of two kinds of monopole sources for the two kinds of vortex lines. Formally we may write the hedgehog creation operator $h^\dagger$ as
\begin{equation}
h^\dagger = m^\dagger_{2+} m_{2-}
\end{equation}
where $m_{2\pm}$ create the two kinds of monopole sources.
Now the charge $1$ of the hedgehog implied by the bulk $\theta$ term implies that these monopole sources are charged. Further as $Z^T_2$ changes the sign of $N_z$, it
interchanges the two monopole sources. It follows that the monopoles $m^\dagger_{\pm}$ each carry charge $ \pm 1/2$ of the $\psi_1$ field. This is precisely what is implied by the Witten effect as applied to the two component BF + FF topological field theory of the previous section.

Exactly the same description can also be provided for the boson topological insulator with symmetry $U(1) \rtimes Z_2^T$. Then we consider anisotropy similar to the above with $\psi_{1,2}$ charged under the global $U(1)$ symmetry but let $\psi_{1,2} \rightarrow \psi_{1,2}$, $N_z \rightarrow - N_z$  under $Z_2^T$.  The rest of the discussion is identical to the one above.

This establishes the connection between the bulk sigma model and topological field theory descriptions. Apart from giving an alternate perspective we hope that the ideas of this section will provide insights on physical realization of these SPT phases, a task we leave for the future.

 \section{Other symmetries: Topological Paramagnets}
 \label{Sec:OtherSymm}
Now let us study various other symmetries which are particularly appropriate to quantum spin systems. By analogy with electronic topological insulators, SPT phases in quantum magnets may be christened topological paramagnets. In particular we highlight two  cases: (i) $Z_2^T$ time reversal symmetry. This is the simplest symmetry that produces a topological phase and we construct a nontrivial phase thus indicating a $Z_2$ class. In the absence of a conserved charge there is no quantized magnetoelectric effect. A separate topological phase with this symmetry, but chiral modes on a domain wall is constructed in the next section.  (ii) $U(1)\times Z^T_2$ This corresponds physically to a time reversal invariant spin system where the $z$ component of spin is conserved. Two nontrivial phases are constructed. The first has a quantized $\theta=2\pi$, but the symmetry prohibits background charge which allows us to sharply define statistics of vortices. The possibility of an exotic type of bose liquid, the vortex metal, as a generic surface state is discussed. The other nontrivial phase has $\theta=0$. However, in this case we show that a deconfined quantum critical action could emerge on tuning just a few parameters. We relegate to the Appendix the following symmetry which is also readily analyzed:  (iii)  $U(1)\rtimes Z_2$  for which we obtain  $Z_2$ topological phases. This is of interest since it does not involve time reversal symmetry.

\subsection{Symmetry $Z_2^T$}
\label{Sec:TimeRev}
We now consider the case of just time reversal symmetry,  both by analyzing the projective representations of surface vortices and by constructing bulk field theories

 {\bf Surface Theory:}
As usual it is convenient to assume a slightly bigger symmetry to identify the relevant physics and then break it down to the physical symmetry. Here it is sufficient to enlarge the symmetry to $U(1)\times Z_2^T$, so we may discuss vortices in a boson field $b_1^\dagger=e^{i\phi_1}$.

Let us first discuss the transformation of the $\phi_1$ field under time reversal. If this was like `charge-phase', then under time reversal $\phi_1\rightarrow -\phi_1+\eta \pi$, where $\eta=0,\,1$.  However, we would be able to pin this for either value of $\eta$ by either a $\cos \phi_1$ or $\sin\phi_1$ term. So this does not correspond to an SPT phase boundary state, since the surface can be gapped in a trivial fashion, without breaking symmetry. The other option is that $\phi_1$ transforms like the XY spin, i.e. $\phi_1\rightarrow \phi_1+\eta \pi$. In this case, for $\eta= 1$, the term that can be added to the Lagrangian is $\cos (2\phi_1+c)$ and $\phi_1$ field cannot be gapped without breaking symmetry. Now we need to consider the transformation of the vortices because that can lead to a gapped state even if $\phi_1$ itself cannot be condensed. Before doing that we note that the field $n_1$ conjugate to the phase $\phi_1$ transforms as $n_1\rightarrow -n_1$ under time reversal (i.e. like $S_z$ spin) to preserve the commutation relations.

 Let us now discuss the transformation properties of the vortices, under the remaining symmetry $Z_2^T$. What are the projective representations of the symmetry group $Z_2^T$? These are essentially the end states of a 1D topological phase with this symmetry. It is well known that there is a Z$_2$ classification of such phases and the nontrivial phase is just the Haldane phase with gapless edge states which are spin $1/2$ objects. Therefore the  projective representation of $Z_2^T$ which we need is the following transformation of two vortex fields (just like spin 1/2)  under time reversal, so $\psi_{2+} \rightarrow +\psi_{2-}$,  $\psi_{2-} \rightarrow -\psi_{2+}$ or more compactly
\begin{eqnarray}
\Psi_2 \rightarrow i\sigma_y \Psi_2
\end{eqnarray}
Note, that if the vortex fields condense, they can either condense individually, or simultaneously. In the former case the gauge invariant operator $N^z=|\psi_{2+}|^2-|\psi_{2-}|^2$ takes on a nonzero expectation value. This object, under time reversal, transforms as $N^z\rightarrow -N^z$, so this breaks time reversal.  If both condense, then the gauge invariant field $\psi_{2+}^*\psi_{2-}=e^{i\phi_2}$ acquires an expectation value. However, under time reversal this also transforms nontrivially $\phi_2\rightarrow \phi_2+\pi$ (note $\phi_1$ and $\phi_2$ transform the same way) and cannot take on an expectation value without breaking time reversal symmetry (both these are spin operators essentially). A third option is that $\psi_{2s}$ do not condense individually, but a pair condenses. This leads to a Z$_2$ topologically ordered state that does not break symmetry, which is also consistent with a topological surface state.

The effective theory for bosons at the edge consistent with this symmetry is
\begin{eqnarray}
{\mathcal L}_e &=& \sum_s|(\partial_\mu -i\alpha_{2\mu})\psi_{2s}|^2 \nonumber\\&&+\frac{1}{2\kappa} f^2_{2\mu\nu}+ \sum_{m}(\lambda_m V^{2m}+ c.c)
\label{NCCP1}
\end{eqnarray}
where $V_{2m}=e^{i2m\phi_1}$ is the $2m$ monopole insertion operator, which is allowed once we break the $U(1)$ symmetry to leave just the time reversal invariance which allows even numbers of monopoles.  Note, the background magnetic field  $(\partial_x\alpha_{2y}-\partial_y\alpha_{2x})=2\pi n_1$, is odd under time reversal symmetry and is not allowed.  Note however  the boson mixing terms $e^{i(\phi_1\pm \phi_2)}$ are allowed by symmetry. Other symmetry allowed terms are the same as with $U(1)\times Z_2^T$ symmetry which is discussed  below Eqn\ref{NCCP1U(1)spin}.

{\bf Surface $Z_2$ Topological Order and Symmetry:} Just as we did in Section \ref{SurfaceAnalysis} it is extremely instructive to consider the question of why a trivial paramagnetic state is not allowed at the surface from the point of view of the $Z_2$ topological surface state. Here of the three non-trivial topological quasiparticles, the two bosonic ones each transform as Kramers doublets under $Z_2^T$, i.e. ${\mathcal T}^2=-1$. They are simply the unpaired vortex of either $b_1$ or $b_2$. They have mutual semionic statistics. Their bound state is a $Z_2^T$ singlet (${\mathcal T}^2=+1$) but it is a fermion. To destroy the topological order we must condense one of these non-trivial quasiparticles. However when either of the two bosonic excitations condense $Z_2^T$ is spontaneously broken. The fermion cannot condense, and time reversal symmetry prohibits nontrivial Chern number for fermions. Thus there is no possibility of a trivial paramagnet.

The cohomology classification \cite{chencoho2011} also produces 
 one nontrivial SPT
phase with this symmetry. Our analysis gives a direct understanding of the allowed surface structure of  this phase. However later we point out another distinct non-trivial SPT phase with $Z_2^T$ symmetry that appears to be beyond the cohomology classification.

{\bf 3D Bulk Lagrangian:}
Consider the bulk Lagrangian:
\begin{eqnarray}\label{BF Lagrangian_new}
\mathcal{L}_{BF} &=& \frac{\epsilon}{2\pi}(B_1\partial a_1+B_2\partial a_2) + \Theta\frac{\epsilon}{4\pi^2}\partial a_1\partial a_2
\end{eqnarray}
the first term is invariant under time reversal if we assume $B_I^{0i}\rightarrow B_I^{0i}$ and $a_I^i\rightarrow a_I^i$ while $B_I^{ij}\rightarrow -B_I^{ij}$ and $a_I^0\rightarrow -a_I^0$ under time reversal. The transformation law for $B$ is obtained by assuming it is connected to a conserved current that  transforms like spin current. Thus, unlike Chern-Simons in 2D, the bulk action is naturally invariant under T. However the second term changes sign, if both $a_I$ transform in the same way. This fixes $\Theta$ to one of two values $0$ and $\pi$  yielding {\em at least} two phases. Since there is no conserved charge there is no coupling to an external field. However, the `fractionalized' degrees of freedom at the ends of vortices is captured in this formalism.

It is relevant to note that phenomena used to distinguish topological phases  previously do not apply for this symmetry. For example, the absence of a conserved charge does not allow us to define the magneto electric polarizability. Also, it turns out that domain wall between opposite time reversal symmetry breaking surfaces do not carry gapless modes. Recall, in free fermion 3D topological insulators and class DIII topological superconductors there are chiral edge modes on T-breaking domain walls. Also, for bosonic topological insulators there was a non chiral but protected domain wall mode. Here the domain wall is nonchiral (as can be seen from the $K=\sigma_x$ matrix that enters in the second term) and in the absence of a conserved charge the oppositely propagating modes can acquire a gap. Nevertheless, the surface states are still special, and are either gapless, break symmetry or develop topological order.

\subsection{Symmetry U(1)$\times\ Z^T_2$}
 Here the U(1) can be interpreted as spin rotation symmetry about $z$ axis. However, we will often continue referring to the conserved quantity as charge. We construct two different topological phases (and the composition of these phases defines a third non-trivial phase), which are interesting for different reasons.

{\bf Surface Theory of Phase 1 and Deconfined Criticality action:}
Here we do not enlarge the symmetry. Let bosons $b^\dagger_1=e^{i\phi_1}$ be charged under the $U(1)$ symmetry so:
\begin{eqnarray}
\phi_1&\rightarrow&\phi_1+\epsilon\,\,:U(1)\nonumber\\
n_1&\rightarrow&n_1\,\,:U(1)\nonumber\\
\phi_1&\rightarrow&\phi_1+\pi\,\,:Z^T_2 \nonumber\\
n_1&\rightarrow&-n_1\, \,:Z^T_2
\label{transformP1}
\end{eqnarray}
Next, we consider the vortices $\Psi_2$ of the field $e^{i\phi_1}$, and specify their transform under the remaining time reversal symmetry, which has a single projective representation:
\begin{equation}
\Psi_2 \rightarrow i\sigma_y \Psi_2 \, \,:Z^T_2
\end{equation}
. Now, since $\psi_{2+}^*\psi_{2-}\sim e^{i\phi_2}$ we have $\phi_2 \rightarrow \phi_2+\pi $ under time reversal.
The effective field theory is written as:

\begin{eqnarray}
{\mathcal L}_e &=& \sum_\sigma |(\partial_\mu -i\alpha_{2\mu})\psi_{2\sigma}|^2 +\frac{1}{2\kappa} f^2_{2\mu\nu}\nonumber \\&&-\lambda [(\psi_{2+}^*\psi_{2-})^2+{\rm h.c.}]+{\mathcal V}(|\Psi_2|^2))
\label{NCCP1U(1)spin}
\end{eqnarray}

The second last term is $\cos2\phi_2$ which is preserves time reversal symmetry. The flux $(\partial_x\alpha_{2y}-\partial_y\alpha_{2x})/2\pi = n_1$ vanishes on  average since the density $n_1$ changes sign under time reversal.  No monopole insertion operators are allowed since changing the flux corresponds to inserting conserved U(1) charge. We note that this action is very similar to the easy plane non compact CP$_1$ (NCCP$_1$) action, proposed as the critical theory between a spin 1/2  easy plane Neel antiferromagnet and  Valence Bond Solid (VBS) order. The flux here is just the spin density, while the vortex bilinear $e^{i\phi_2}$ correspond to the VBS order. In contrast to the square lattice with four fold rotation symmetry, here the square of the VBS order parameter is allowed, as on a rectangular lattice. An important distinction from previously discussed deconfined criticality is that here translation symmetry is not invoked.

Symmetry actually permits other terms in this action, for example, the linear derivative terms $(\partial_t\alpha_{2i}-\partial_i\alpha_{20})$, which corresponds to electric fields (spin currents) in the ground state. Similarly, finite gauge charge is also allowed in the ground state, corresponding to finite vortex density, since vortices here do not break time reversal symmetry. This introduces linear time derivative terms in the action above. However, if we expand the symmetry to include a $Z_2$ that reverses the orientation of the U(1) rotation axis, i.e. that sends $n_1\rightarrow -n_1$ (which is a rotation by $\pi$ around the $S_x$ axis in spin notation), then this prohibits the additional terms discussed here, since both electric field and gauge charge are odd under this Z$_2$. Thus, for the topological surface state with symmetry $[U(1)\rtimes Z_2] \times Z_2^T$ the field theory is given by \ref{NCCP1U(1)spin}. Parenthetically we note that precisely this internal symmetry was also assumed in the original discussion of deconfined crticality in $2d$ quantum magnets with easy plane anisotropy\cite{deccp}. Generically, either the bosons or the vortices are condensed, which implies that either $U(1)$ or time reversal symmetry is broken. However, if the critical point separating these states is stable to fluctuations, then one could tune a single parameter and access a deconfined critical point on the surface. It is presently unclear if this is true for the theory in  Eqn. \ref{NCCP1U(1)spin}, which is an easy plane NCCP$_1$ with a two fold $\lambda$ anisotropy term. There is mounting evidence that the SU(2) symmetric NCCP$_1$ model supports a stable quantum critical point. While initial studies were divided between continuous \cite{lesikav2004,MotrunichAv} and weak first order\cite{Prokofiev,Wiese}, recent studies of quantum models seem to favor continuous transition \cite{Sandvik, Damle, Kaul}. However the situation is less clear with easy plane anisotropy\cite{lesikav2004,Prokofiev2}, and the $\lambda$ anisotropy term above.  The connection to SPT surface states should provide additional motivation for further study.

Thus far we have assumed translation invariance on the surface, but in fact only internal symmetries are required to define the phase. The presence of surface randomness that respect internal symmetries will provide random variations in the local critical coupling. This random energy density term is known to be typically relevant at a quantum critical point \cite{Chayes}, since it requires a rather stringent condition to be met, $\nu>1$ for irrelevance in a clean critical pint in $d=2$. Here we emphasize a crucial difference with the realization of the deconfined critical theory in $2d$ quantum magnets. In that case the presence of disorder leads to a random field that couples linearly to the VBS order parameter. In the spinon representation this is a random monopole insertion term. Alternately in the dual vortex representation this is a random term that couples to $\psi_{2+}^*\psi_{2-}$. This coupling is expected to be relevant at the clean deconfined critical point and might potentially lead to confinement at the resulting disordered fixed point. Thus it is not clear if the NCCP$_1$ description is a useful one in the presence of disorder. In the present problem however a linear coupling to $\psi_{2+}^*\psi_{2-}$ remains forbidden even in the presence of disorder. The random energy terms though relevant are still not expected to lead to confinement by themselves. More dangerous potentially are random variations in the coupling $\lambda$.  The fate of the disordered NCCP$_1$ model in the presence of this particular kind of randomness remains to be investigated.  In this context it may be relevant to note that even the fate of the 3D fermonic topological insulator surface states in the presence of disorder and interactions is also not a settled issue. It is currently unclear if one of the symmetries is spontaneously broken in the low energy limit. If the symmetries are preserved, a critical metal with universal conductance is predicted\cite{Mirlin}.

As in the other examples a surface state with $Z_2$ topological order that preserves all symmetries is allowed and is readily accessed by condensing paired vortices. For both
Phase 1 and Phase 2 discussed in this subsection with symmetry $U(1) \times Z_2^T$, the symmetry properties of the corresponding surface topological order is summarized in Appendix \ref{App:Z2}.

{\bf Surface Theory of Phase 2 and Vortex Spin Metal:}
A different topological phase is accessed by enlarging the symmetry momentarily to $[U(1)\times U(1)] \times Z_2^T$. Now, we can assume that boson of species 1, $b^\dagger_1=e^{i\phi_1}$ be charged under the first $U(1)$ symmetry and transforms exactly as in Eqn.\ref{transformP1}. Vortices in this boson field transform under the remaining $U(1)$ and time reversal symmetry as:

\begin{eqnarray}
\Psi_ 2\rightarrow e^{i\epsilon' \sigma_z/2}\Psi_2 \,\,&&:U(1)\\
\Psi_2 \rightarrow  i\sigma_y \Psi_2 \,\,&&:Z^T_2
\label{P2}
\end{eqnarray}

Now, the effective theory at the surface is:
\begin{equation}
{\mathcal L}_e = \sum_\sigma |(\partial_\mu -i\alpha_{2\mu})\psi_{2\sigma}|^2 +\frac{1}{2\kappa} f^2_{2\mu\nu}\nonumber +{\mathcal V}(|\Psi_2|^2))
\label{NCCP1U(1)spinP2}
\end{equation}

which has neither monopole insertion nor anisotropy terms due to the presence of separate boson species conservation. There is no background flux due to time reversal symmetry.  However, breaking this down to the single $U(1)$ symmetry one is allowed the  following term allowed $\cos(\phi_1-\phi_2)$ since both bosonic fields transform the same way under symmetry. Now $V^*_1=e^{i\phi_1}$ corresponds to a monopole insertion operator - thus this is a composite operator  which in the variables above may be written as:  $\psi_{2+}^*\psi_{2-}V_1 + {\rm h.c.} $. This breaks down the $U(1)\times U(1)$ symmetry to a single $U(1)$, and leads to a binding of their vortices. Now, if only the time reversal symmetry is broken, then a quantized Hall effect results of the conserved spin. The discussion closely parallels that in Section \ref{Sec:bosonicTI}.

However, if U(1) symmetry is broken, the surface is an $XY$ ordered state of the spin system, and the vortices can be shown to be fermionic as in Section \ref{SurfaceAnalysis}. An advantage in the topological paramagnet  compared to the bosonic topological insulator surface is the absence of background spin density that implies the fermionic vortices move in zero background field. This allows for a sharp definition of their statistics in terms of the Berry phase under exchange. Moreover, since the vortex density does not break time reversal symmetry, generically a finite vortex density will be present in the ground state. In the $XY$ ordered state these vortices will form a vortex solid and their statistics is not very important. It is however extremely interesting to ask about the result of destroying the $XY$ order by melting the vortex solid and  proliferating the vortices. With Fermi statistics the vortices will form a Fermi surface which will be coupled to the non-compact $U(1)$ gauge field. The resulting state is a ``vortex spin metal" - a compressible metallic phase of spins with many interesting properties.  It is a gapless spin liquid with a vortex Fermi surface and is distinct from the more familiar 2d quantum spin liquids with a spinon Fermi surface. Ref. \onlinecite{Galitski} proposed a very analogous  vortex metal phase as an exotic possibility for a magnetic field driven quantum vortex liquid state in two space dimensions. There the magnetic field explicitly breaks time reversal invariance. In contrast  the vortex spin metal obtained at the surface of the 3d topological paramagnet is a phase that preserves the defining $U(1) \times Z_2^T$ symmetry. As with the other examples discussed in this paper such a time reversal invariant vortex spin metal is presumably forbidden in strict 2d spin systems.

{\bf 3D Bulk Theory} For both Phase 1 and Phase 2, the 3D topological theories are identical, and only differ in the coupling of the conserved charge  to the external field:
\begin{eqnarray}
{\mathcal L}_{tot}&=&{\mathcal L}_{topo}+{\mathcal L}_{em}\\
{\mathcal L}_{topo} &=& \frac{1}{2\pi}\epsilon B_I\partial a_I + \Theta\frac{\epsilon}{4\pi^2}\partial a_1\partial a_2\\ \nonumber
\end{eqnarray}
Under the Z$^T_2$ symmetry, $a_{1i} \rightarrow a_1$, and $a_{2i} \rightarrow a_{2i}$ while their $0$ components change sign. Thus  the `axion' field $\Theta$ must be odd under Z$_2$ so the action as a whole is invariant. This allows us to fix $\Theta=0,\,\pi$, and of course we pick the latter value in the topological phase. In general, in a 3D topological phase protected by time reversal, both fields should transform in the same way.

Now, Phase 1 has a single charged bosons, $\phi_1$,
\begin{equation}
{\mathcal L}^{Phase 1}_{em} = \frac1{2\pi}\epsilon B_1\partial A
\end{equation}
 The bulk theory predicts $\theta=0$, i.e. no magnetoelectric effect for this phase.

 However, for Phase 2, both bosons are charged so:
 \begin{equation}
{\mathcal L}^{Phase 2}_{em} = \frac1{2\pi}\epsilon(B_1+B_2)\partial A
\end{equation}
  and the bulk theory predicts $\theta=2\pi$ magnetoelectric effect for this phase.

\section{3D E$_8$ phase with Half Quantized Surface Thermal Hall Effect and Miscellaneous Comments}
\label{Sec:E8}
Thus far we have based our discussion of novel 3D SPT phases on the two dimensional $K=\sigma_x$ matrix. When a conserved charge is present, these phases often lead to a quantized magnetoelectric effect, or equivalently, a half quantized surface Hall effect. On general grounds one may expect additional phases based on the fact that thermal transport can also be quantized.  In these phases chiral modes are expected at the domain walls between opposite symmetry breaking regions, that lead to the quantized thermal Hall conductance. In this section we provide a possible field theoretic description of such a phase.

Recall that in a 2D system, the combination $\kappa_{xy}/T=\nu_T \frac{\pi^2k_B^2}{3h}$ is quantized, where $\kappa_{xy}$ is the thermal Hall conductance and $T$ is temperature, in the limit of $T\rightarrow 0$. Here $\nu_T$ counts the number of chiral boson modes at the edge. For bosons with SRE in $d=2$, it is known that the quantization  takes values $\nu_T=8n$ that are multiples of 8 times quantum of thermal conductance. Anything else leads to topological order. These states are based on the $K$ matrix of the Kitaev $E_8$ state\cite{luav2012}:

\begin{equation}
K^{E_8}=\left(
    \begin{array}{cccccccc}
      2 & -1 & 0 & 0 & 0 & 0 & 0 & 0\\
      -1 & 2 & -1 & 0 & 0 & 0 & 0 & 0 \\
      0 & -1 & 2 & -1 & 0 & 0 & 0 & -1 \\
      0 & 0 & -1 & 2 & -1 & 0 & 0 & 0 \\
      0 & 0 & 0 & -1 & 2 & -1 & 0 & 0 \\
      0 & 0 & 0 & 0 & -1 & 2 & -1 & 0 \\
      0 & 0 & 0 & 0 & 0 & -1 & 2 & 0 \\
      0 & 0 & -1 & 0 & 0 & 0 & 0 & 2 \\
    \end{array}
  \right)
  \label{E8}
\end{equation}

One may utilize this fact to construct the following 3D SPT phase. Assume time reversal symmetry $Z_2^T$ is present . Consider the three dimensional theory given by
\begin{equation}
{\mathcal L}= \frac1{2\pi}\sum_{I=1}^8 \epsilon B_I\partial a_I + \Theta \sum_{I,J}\frac{K^{E_8}_{IJ}}{8\pi^2}\epsilon \partial a_I \partial a_J
\end{equation}
As long as all the fields $a_I$ transform the same way under time reversal symmetry, the coefficient $\Theta$ is quantized to $\Theta=0,\, \pi$. The latter leads to a topological phase. If time reversal symmetry is preserved in the bulk but broken on the surface, it is readily seen that each domain has thermal Hall conductivity $\nu_T=\pm 4$, and a domain wall between opposite domains has the eight chiral edge states of the two dimensional theory specified by the $K$ matrix in Eqn. \ref{E8}.

It is likely that such a state lies beyond cohomology classification since we have already identified a phase based on $K=\sigma_x$ in this symmetry class that exhausts the set of states predicted by cohomology theory\cite{chencoho2011}. A  question that is relevant in this context is whether the field theory above can be realized within a lattice model. One  difference from the other topological phases we have described based on $K=\sigma_x$  `FF' term is that in those cases a lattice regularization of the field theory can be readily envisaged since it involves a Berry phase for the product of electric and magnetic fields representing two different species of vortices $\epsilon F_1F_2\rightarrow E_1\dot B_2 +E_2\cdot B_1$. This term is naturally discretized by assuming the corresponding vector potentials live on the links of the direct lattice and the dual lattice. However, the diagonal entries of the $K^{E_8}$ matrix above lead to terms that are not obviously compatible with a lattice. Whether this imposes an additional constraint on possible phases is an important open question. If indeed the phase described above is physically admissible, then
it remains to be clarified if the additional states lead to a $Z_2$ or a $Z$  extension (assuming just time reversal symmetry). We leave these questions to future study. In this context it may be relevant to note that the analogous free fermion phases are topological superconductors in 3D protected by time reversal symmetry (Class DIII), which are classified by integers.  On the other hand, one may ask what are the properties of the topologically ordered surface state that is fully symmetric. We conjecture that a candidate state is a Z$_2$ topological ordered state where all three nontrivial excitations are fermionic, and have $\pi$ mutual statistics. Such a state when realized in 2D is given by the $K$ matrix:
 \begin{equation}
K^{SO(8)}=\left(
    \begin{array}{cccccccc}
      2 & -1 & -1 & -1 \\
      -1 & 2 & 0 & 0\\
      -1 & 0 & 2 & 0 \\
      -1 & 0 & 0 & 2 \\
    \end{array}
  \right)
  \label{SO(8)}
\end{equation}
which is the Cartan matrix of SO(8). This state when realized in 2D has 4 chiral edge states, and hence must break time reversal symmetry. However, it may appear on the surface of a 3D SPT phase, protected by time reversal symmetry.

\section{Conclusions}
In summary, we would like to highlight the remarkable similarities between free fermion topological insulators, and the bosonic interacting topological phases described here.
In the former case, the surface is gapped only on breaking one of the defining symmetries of charge conservation or time reversal symmetry. Then, the resulting ordered phase also possesses unusual properties, for example when charge conservation is destroyed by a superconducting surface, the vortices carry a Majorana zero mode. Similarly, for the bosonic topological insulator with the same symmetries, breaking charge conservation at the surface leads to fermonic vortices (albeit without an attached Majorana zero mode). On the other hand, breaking just time reversal symmetry leads for the fermionic case to a quantized magneto electric effect of $\theta=\pi$, whereas for bosonic TIs in the same situation, the same response is quantized, but at $\theta=2\pi$. The fully symmetric surface of the fermonic TI, from which these conclusions can be readily derived, is a Dirac dispersion of free fermions. We propose that the analog for bosons is the deconfined quantum critical action, which describes a putative gapless state from which, on being subject to various perturbations, realizes  different ground states of the surface. It is also relevant to note that bosonic analogs of topological superconductors exist where domain walls between opposite time reversal symmetry breaking regions carry gapless chiral modes.

It is interesting to further highlight the particular case of spin systems. The spin analogs of topological insulators - the topological paramagnets - may potentially be the most important realization in solid state systems of the class of phases we have described. The surface of the topological paramagnet either spontaneously breaks symmetry or is in a
quantum spin liquid state that is not allowed to exist in strict two dimensions with the same symmetry. We have discussed examples of such quantum spin liquids with surface topological order or with exotic gapless excitations. Our work raises fascinating questions on what kinds of spin liquids with symmetry are actually allowed to exist in strictly 2d systems that should be of direct importance to studies of 2d quantum magnetism.

There are several open questions for future work. Clearly, a central question is whether there are microscopic models, or perhaps even experimentally relevant systems, which could realize these phases.  The bulk sigma model field theories may provide useful guidance in searching for such realizations. One route to accessing SPT phases in two dimensions is to start
with a fractionalized phase and confine
the fractionalized excitations. Our analysis suggests that a similar
route may also be possible in three dimensions by starting with a
fractionalized phase with emergent deconfined $U(1) \times U(1)$ gauge
fields if these are confined by condensation\cite{mpaf} of mutual
dyons (where a monopole of one $U(1)$ gauge field is bound to
particles that carry gauge charge of the other gauge field). Exploring
this possibility might also suggest physical realizations of the 3d
SPT phases. A more formal question is whether one can push the field theoretic descriptions of this paper to obtain all possible SPT phases in 3D, which could shed light on the way in which the chiral phases augment the cohomology characterization. The three dimensional `BF+FF' theories seem a convenient tool to capturing bosonic SPT phases. However, general constraints on the form of such theories is presently unclear.  

\section{Acknowledgements}
A.V. would like to thank Ari Turner and especially Yuan Ming Lu for stimulating discussions and collaborations on related topics, and acknowledges support from NSF DMR-1206728 . TS likewise thanks Liang Fu, Michael Levin, Chong Wang, Z. Gu and Xiao-gang Wen. TS was supported by NSF DMR-1005434. We both thank Matthew Fisher for scintillating discussions and encouragement.  This material is based upon work supported in part by the National Science Foundation under Grant No. PHYS-1066293 and the hospitality of the Aspen Center for Physics. We also thank the  Perimeter Institute for Theoretical Physics, and the Kavli Institute for Theoretical Physics where parts of this work were done. This work was partially supported by the Simons Foundation by award numbers 229736 (TS) and 231377(AV). On completing this work we became aware of other studies which have some overlap with the present work \cite{Swingle},\cite{Cenke}.

\appendix
\section{3D BF Theory: Surface States and EM response:}
\label{appendixBFsurface}
As a warmup let us recall the derivation of the edge states of a Chern Simons theory in 2D\cite{Wenbook}. We specialize to the $K=\sigma_x$  Chern Simons Theory:

\begin{eqnarray}\label{CS Lagrangian}
\mathcal{L}_{CS} &=& \frac{\epsilon^{\mu\nu\lambda}}{2\pi}a_\mu^1\partial_\nu a^2_\lambda
\end{eqnarray}
Note, gauge invariance at the surface can be ensured by working in the gauge $a_0=0$. This implies $d a^I = 0$ so $a^I_i=\partial_i\phi_I$. This gives the edge Lagrangian:
\begin{equation}
{\mathcal L}=\frac1{2\pi}\partial_x\phi_1\partial_\tau \phi_2
\label{commutation}
\end{equation}
leading to the usual Kac-Moody commutation relations.

The edge dynamics originates from other terms. For example, we can add a Maxwell term to the original action $(\partial_\mu a^I_\nu-\partial_\nu a^I_\mu)^2$. The only low derivative term that will appear at the edge is from $\partial_y a^I_x$. Substituting the edge field $a^I_x=\partial_x\phi^I$ and noticing that the derivative perpendicular to the edge (i.e. along $y$), only picks up the confining wavefunction of the edge states we are lead to: $\partial_ya^I_x\propto \partial_x\phi^I$. This gives potential terms
$${\mathcal L}_1 = \rho[(\partial_x\phi_1)^2+(\partial_x\phi_2)^2]$$

Note, the pair of fields $\phi_1,\,\phi_2$ which are canonically conjugate Eqn. \ref{commutation} is like any regular one dimensional Luttinger liquid. The special physics of SPT phases arises from the fact that the fields can transform under the symmetry in ways that a 1D system cannot. For example, in the $U(1)$ protected integer quantum Hall phase of bosons, the transformation law of the first nontrivial phase is: $\phi_i\rightarrow \phi_i+\epsilon$. {\em i.e} both fields transform under the charge rotation.  This leads to protected edge states. By analogy, it appears that we should find that the surface of a 3+1D SPT phase of bosons is a regular two dimensional bosonic system, apart from application of symmetries. Indeed, we show below that this is the surface state of the $3+1$ D $BF$ theory.

Let us begin with the following 3D Lagrangian:
\begin{eqnarray}\label{BF Lagrangian}
\mathcal{L}_{BF} &=& \frac{\epsilon^{\mu\nu\lambda\sigma}}{2\pi}B_{\mu\nu}\partial_\lambda  a_\sigma
\end{eqnarray}

Here, a bosonic current has been written as $j^\mu= \epsilon^{\mu\nu\lambda\sigma}\partial_\nu B_{\lambda\sigma}/2\pi$, and $\partial\wedge a$ represents the vortex loops.

To derive surface properties, again, the non-dynamical parts of the Lagrangian implements the constraint: $\epsilon_{ij}\partial_ia_j=0$ and $\epsilon_{ijk}\partial_iB_{jk}=0$. One can solve this to obtain $a_i=\partial_i\phi$ and
$B_{ij}=\epsilon^{ij}\partial_i \alpha_j$. We take the gauge $B_{0i}=a_0=0$. The topological part of the edge Lagrangian
(the edge is taken to be perpendicular to z and we use the indices $a,b=(x,\,y)$):

\begin{equation}
{\mathcal L}=\frac{\epsilon^{ab}}{2\pi}\partial_a\alpha_b{\partial_\tau \phi}
\end{equation}

Now, for the dynamics, once again one introduces 'Maxwell' terms in the bulk:
$(da)^2$ and $(dB)^2$. Again, the ones that survive with low derivatives have $\partial_z$
acting on them. The three terms that appear are:

\begin{equation}
{\mathcal L}_1 = \rho_1[(\partial_x\phi)^2+(\partial_y\phi)^2]+\rho_2[(\partial_x\alpha_y-\partial_y\alpha_x)^2]
\end{equation}
. Thus
our boundary Lagrangian is (${\mathcal S} =\int dxdyd\tau\,{\mathcal L}_e$ ):

\begin{eqnarray}
\label{BF Lagrangian1}
{\mathcal L}_e&=&\frac{\epsilon^{ab}}{2\pi}\partial_a\alpha_b\partial_\tau{\phi}\\ \nonumber
&& +\rho_1[(\partial_x\phi)^2+(\partial_y\phi)^2]+\rho_2[(\partial_x\alpha_y-\partial_y\alpha_x)^2]
\end{eqnarray}

One interpretation of this Lagrangian is that of a photon on a 2D surface  where the Gauss law constraint has been solved i.e. $2+1$ D, the Gauss law $\partial_xE_x+\partial_yE_y=0$ can be solved by writing $E_a=\epsilon_{ab}\partial_b\phi/\pi$. Then, the term that leads to canonical quantization of electric fields: $E_a\partial_\tau \alpha_a$ is now replaced by $\frac{1}\pi \partial_\tau\phi(\partial_x\alpha_y-\partial_y\alpha_x)$, the first term in ${\mathcal L}_e$. The Hamiltonian of the Maxwell theory: $\rho_1[E^2_x+E^2_y]+\rho_2 (\partial_x\alpha_y-\partial_y\alpha_x)^2$ is the term written above in the Lagrangian.

Although Refs. \onlinecite{Aratyn,ChoMoore} follow a rather similar derivation, they interpret the theory above as the bosonized description of a 2+1D Dirac fermion. This interesting speculation does not appear to be compatible with the well known fact that the theory described by Eqn.\ref{BF Lagrangian1} is dual to a 2+1D theory of bosons. Explicitly this can be seen as follows. Since $\epsilon^{ab}\partial_a\alpha_b/2\pi$  is conjugate to the phase $\phi$, we denote it by $\Pi_\phi=\epsilon^{ab}\partial_a\alpha_b/2\pi$ and use this field to write the Hamiltonian of the surface theory as:

\begin{equation}
H={4\pi^2\rho_2}\Pi_\phi^2 + \rho_1 (\nabla\phi)^2
\label{Harmonic}
\end{equation}
This is just the theory of a boson in the two spatial dimensions of the surface (as expected, since we began with a bosonic theory).

It is useful to catalog the connection between the dual descriptions. The monopole insertion operator in the surface electrodynamics is $e^{i\phi}$ and actually corresponds to the insertion of particles. The other excitations, `gauge charges' of the gauge theory, are the ends of vortices of the 3D bulk and are point particles on the surface. Here they behave like charges in the 2D electrodynamics, since vortices of the $\phi$ field are equivalent to violating Gauss law for the electric field. Also, since ${\rm Curl} a\rightarrow {\rm Curl} \nabla\phi$ they correspond to the ends of the three dimensional vortices. The vortex insertion operator is of course a nonlocal object, which reflects the fact that one cannot insert a gauge charged particle in the bulk without changing the gauge fields everywhere. The 2D surface is gapped either by monopoles $e^{i\phi}$ or by vortex condensation (Higgs mechanism).  However symmetry may forbid these leading to SPT phases.

Let us briefly review some questions that arise in the context of the Lagrangian (\ref{BF Lagrangian}). One can add terms such as ${\mathcal L}_1 \sim (\epsilon \partial B)^2$ and ${\mathcal L}_2 \sim (\partial_\mu a_\nu -\partial_\mu a_\nu)^2$ which are local and respect symmetries. Integrating out $B$ now appears to give the `Higgs' term $a_\mu^2$. However, it is readily seen that this still describes an insulator, by coupling the charge to an external electromagnetic potential $A$ via $\epsilon A\partial B/2\pi$. Now, integrating the fields $B$ essentially enforces $a\sim A$, which when substituted into ${\mathcal L}_2$  simply produces a Maxwell action for the external field: $\sim (\partial_\nu A_\mu-\partial_\mu A_\nu)^2$ as expected for an insulator. A 3D topological EM response will appear in other cases where an `FF' term is present, by the same substitution.

\section{$\Theta$ periodicity in multicomponent BF theory}
\label{Kinteger}
In this Appendix we prove the $2\pi$ periodicity of $\Theta$ for the multicomponent BF theory. The Lagrangian is
\begin{equation}
{\cal L} = \frac{1}{2\pi}a_\mu^I \epsilon_{\mu\nu\lambda\kappa} \partial_\nu B^I_{\lambda \kappa} + \frac{\Theta}{8\pi^2} K^{IJ} \epsilon_{\mu\nu\lambda\kappa} \partial_\mu a^I_\nu \partial_\lambda a^J_\kappa
\end{equation}
Summation over repeated component indices $I,J,...$ is implicit.
The crucial second term when expressed in terms of the electric fields $\vec e^I$ and the magnetic fields $\vec b^I$ takes the form
\begin{equation}
\frac{\Theta}{8\pi^2}\sum_I K^{II} \left [\left(2 \vec e^I. \vec b^I \right) + \sum_{J>I}K^{IJ} 2\left(\vec e^I. \vec b^J + \vec e^J. \vec b^I \right) \right ]
\end{equation}
Consider the theory on a closed three manifold such as a 3-torus of size $L \times L \times L$. Through one cycle, say the xy cycle, slowly insert $2\pi n_I$  magnetic flux of species $I$ at a rate $\frac{d\Phi_I}{dt}$. This leads to a bulk electric field along the $z$-direction:
\begin{equation}
e^I_z = \frac{1}{L} \frac{d\phi_I}{dt}
\end{equation}
Next slowly turn on $2\pi m_I$ flux of $b^I_z$ in the bulk so that
\begin{equation}
b^I_z = \frac{2\pi m_I}{L^2}
\end{equation}
The quantum amplitude for these processes is given by the $\Theta$ term in the action and takes the form
\begin{eqnarray}
& & e^{i\frac{\Theta}{4\pi^2} \int dt L^3 \left(\frac{2\pi}{L^2}\right)\left[\sum_I\left ( K^{II} m_I \frac{d\Phi_I}{dt} + \sum_{J> I} K^{IJ} m_J \frac{d\Phi_I}{dt} \right)\right ]} \nonumber\\
& = & e^{i\Theta \sum_I \left (K^{II} n_I m_I + \sum_{I \neq J} K^{IJ} n_I m_J \right)}
\end{eqnarray}
For some particular pair $I, J$ choose $n_I = 1, m_J = 1$, and all other $n_{I'} = m_{J'} = 0$. Then the amplitude simply becomes $e^{i\Theta K^{IJ}}$. If all the elements $K^{IJ}$ are integers it follows that $\Theta$ is periodic under a $2\pi$ shift.

\section{Other Symmetries: U(1)$\rtimes\ Z_2$}

Here the U(1) can be interpreted as spin rotation symmetry about $z$ axis, while the Z$_2$ is spin rotn. by 180 degrees about the $x$ axis.

 {\bf Surface Theory:} Let bosons $b^\dagger_1=e^{i\phi_1}$ be charged under the $U(1)$ symmetry. Then the phase $\phi_1$ and conjugate number $n_1$ transform as:
\begin{eqnarray}
\phi_1&\rightarrow&\phi_1+\epsilon\,\,:U(1)\nonumber\\
n_1&\rightarrow&n_1\,\,:U(1)\nonumber\\
\phi_1&\rightarrow&-\phi_1\,\,:Z_2 \nonumber\\
n_1&\rightarrow&-n_1\, \,:Z_2
\label{transformP3}
\end{eqnarray}

Now we would like to understand how a vortex in a superfluid surface state of this boson field will transform. The remaining $Z_2$ symmetry will act on the vortices, however, it is readily seen this switches vortices to anti-vortices. More formally, the vortex fields $\Psi_2$ are coupled minimally to the gauge field $\alpha_2$ whose flux is the number density $n_1=(\partial_x\alpha_{2y}-\partial_y\alpha_{2x})/2\pi$. Now since the number density changes sign under Z$_2$, so does the gauge field $\alpha_2\rightarrow -\alpha_2$. This implies that for the minimal coupling to remain invariant, we need $\Psi_2 \rightarrow \Psi_2^*$. In fact the desired transformation is:

\begin{equation}
\Psi_2 \rightarrow i\sigma_y \Psi_2^* \, \,:Z_2
\end{equation}

This may be viewed as the single projective representation of $U(1)\rtimes Z_2$, where the $U(1)$ may be viewed as the gauge $U(1)$ that changes sign under Z$_2$. It is readily verified that the gauge invariant combinations $|\psi_{2+}|^2-|\psi_{2-}|^2$ and $\psi^*_{2+}\psi_{2-}=e^{i\phi_2}$ both transform nontrivially under  Z$_2$ verifying that if vortices condense they always break the symmetry. A second species of bosons is defined by $b^\dagger_2=e^{i\phi_2}$, which transforms as:

\begin{eqnarray}
\phi_2&\rightarrow&\phi_2+\pi \, \,:Z_2\nonumber \\
n_2&\rightarrow&n_2 \, \,:Z_2
\end{eqnarray}

and is neutral under the global $U(1)$. This satisfies the intuitive requirements of a topological surface state and therefore we conclude a U(1)$\rtimes$Z$_2$ symmetry group also leads to Z$_2$ topological phases. Note however, of the two symmetries were in direct product, there would be {\em no} topological phases.

Let us write down the field theory for the surface in terms of vortices of $\Psi_2$. They are minimally coupled to a vector potential $\alpha_2$ whose flux is the boson density $\nabla \times \alpha_1=n_1$.

\begin{equation}
{\mathcal L}=|(\partial_\mu-i\alpha_{2\mu})\Psi_2|^2 +\frac{1}{2\kappa} f^2_{2\mu\nu}\nonumber +{\mathcal V}(|\Psi_2|^2))
\end{equation}
Since the field $b_1$ is charged, monopole insertion operators are forbidden, but various anisotropy terms, involving four vortex fields are allowed. These and other allowed perturbations are readily identified given the symmetry transformations above.

A dual description of the same theory is obtained by 'fractionalizing' the boson field $b^\dagger_1=\psi^*_{1+}\psi_{1-}$, where $\Psi_1=(\psi_{1+},\,\psi_{1-})$ may be viewed either\cite{deccp} as a Schwinger boson representation of $b_1$, or as vortices of $b_2$.  Now, these transform under a projective representation of the global symmetry $U(1)\rtimes Z_2$:
\begin{eqnarray}
\Psi_1 \rightarrow e^{i\epsilon \sigma_z/2}\Psi_1 \,\,:U(1)\\
\Psi_1 \rightarrow  \sigma_x \Psi_1 \,\,:Z_2
\end{eqnarray}
this is compatible with the transformations in Eqn. \ref{transformP3}. We see this implies that vortices in $b_2$ carry half unit of global charge at the surface. This will help us fix the bulk field theory.

{\bf 3D Bulk Theory} Given the characterization of the surface states above, we can write down a bulk 3D theory that reproduces these features. We write down the following theory based on $K=\sigma_x$ where the conserved charge is coupled to an external electromagnetic field $A$, and later justify it:

\begin{eqnarray}
{\mathcal L}_{tot}&=&{\mathcal L}_{topo}+{\mathcal L}_{em}\\
{\mathcal L}_{topo} &=& \frac{1}{2\pi}\epsilon (B_1\partial a_1+B_2\partial a_2) + \Theta\frac{\epsilon}{4\pi^2}\partial a_1\partial a_2\\ \nonumber
{\mathcal L}_{em} &=&\frac1{2\pi}\epsilon(B_1)\partial A
\end{eqnarray}
Under the Z$_2$ symmetry, $B_1\rightarrow -B_1,\,a_1 \rightarrow -a_1$, but $B_2\rightarrow B_2,\, a_2 \rightarrow a_2$ . Thus  the `axion' field $\Theta$ is odd under Z$_2$ so the action as a whole is invariant. This allows us to fix $\Theta=0,\,\pi$, the latter value yields the topological phase. Also, only one of the boson species carries global U(1) charge which implies that there is no topological contribution to the magnetoelectric polarizability i.e. $\theta=0$.

\section{Symmetry Transformation of Surface States with Topological Order}
\label{App:Z2}
 For convenience we accumulate in this Appendix the properties of the surface state with $Z_2$ topological order of the SPT phases with various symmetries. As described in the main paper such a surface topologically ordered phase provides a particularly simple perspective on why a trivial gapped symmetry preserving surface is not allowed.
The $Z_2$ topological order has four distinct quasiparticles which we will denote $1, e, m, f$. The trivial quasiparticle sector is described by $1$ and consists of all local operators. We will take $e$ (for `electric') and $m$ (for `magnetic') to be bosons and $f$ to be a fermion. $e, m$ and $f$ are all mutual semions. Below we summarize how the physical symmetry is realized for each of the three non-trivial quasiparticles for the various phases. In what follows $q$ demotes the charge under the global $U(1)$ symmetry.  In all cases other possible quasiparticles in the same sector is obtained by adding trivial quasiparticles. The symmetry properties of $f$ follow from those of $e$ and $m$ (as it is a bound state of $e$ and $m$).

\begin{enumerate}
\item
{\bf Symmetry $U(1) \rtimes Z_2^T$}
Here both $e$ and $m$ carry charge-$1/2$ and are time reversal invariant. It is only important to ask about the presence/absence of Kramers degeneracy under $Z_2^T$ corresponding to ${\cal T}^2 = \pm 1$.

 \begin{table}[htdp]
\begin{tabular}{c|c|c}
Field & $q$  & ${\cal T}^2$ \\ \hline
$e$ & $\frac{1}{2}$ & $1$ \\ \hline
$m$ & $ \frac{1}{2}$ & $1$ \\
\end{tabular}
\caption{  $U(1) \rtimes Z_2^T$}
\label{stou1rz2t}
\end{table}%

\item {\bf Symmetry  $Z_2^T$}

Here there are a pair of $e$ particles (denoted $e_\alpha = (e_\uparrow, e_\downarrow)$, a pair of $m$ particles ($m_\alpha = (m_\uparrow, m_\downarrow)$) and a single $f$ particle.  We note that in the absence of other symmetries $e_\uparrow$ will be able to mix with $e_\downarrow^*$ so that we may regard them as a single particle and its antiparticle which together form a Kramers doublet. The same also applies to the $m$ particles.  Therefore we will just work with a single $e$ and single $m$ particle.

 \begin{table}[htdp]
\begin{tabular}{c|c}
Field &    ${\cal T}^2$ \\ \hline
$e$ & $-1 $\\ \hline
$m$ & $-1$ \\
\end{tabular}
\caption{$Z_2^T$ }
\label{stoz2t}
\end{table}%

\item{\bf Symmetry $U(1) \times Z_2^T$}

Here we discussed two phases (Phase 1 and Phase 2): their symmetries are described in Tables
\ref{stou1z2tp1} and \ref{stou1z2tp2}.

 \begin{table}[htdp]
\begin{tabular}{c|c|c}
Field & $q$  & ${\cal T}^2$ \\ \hline
$e$ & $\frac{1}{2}$ & $- 1 $ \\ \hline
$m $ & $0$ & $-1$ \\
\end{tabular}
\caption{ $U(1) \times Z_2^T$: Phase 1}
\label{stou1z2tp1}
\end{table}%

  \begin{table}[htdp]
\begin{tabular}{c|c|c}
Field & $q$  & ${\cal T}^2$ \\ \hline
$e$ & $\frac{1}{2}$ & $- 1 $ \\ \hline
$m$ & $\frac{1}{2}$ & $-1$ \\
\end{tabular}
\caption{$U(1) \times Z_2^T$: Phase 2 }
\label{stou1z2tp2}
\end{table}%

\end{enumerate}

As we emphasized in the main paper the realization of symmetry is such that these $Z_2$ topologically ordered phases cannot arise in strict two dimensional models with local action of the symmetry group.
This is readily seen from the general $K$-matrix classification of $2d$ time reversal invariant gapped abelian phases in Ref. \onlinecite{LevinStern}.  Here we will present an elementary analysis that is sufficient for the purposes of this paper.  Strictly two dimensional systems admit an edge to the vacuum (or equivalently a trivial gapped insulator). without changing the symmetry. This is a key difference from the surface topological order of interest to us here where a domain wall with a trivial gapped insulator is not possible without breaking symmetry. The analysis below, following Ref. \onlinecite{LevinStern}, relies crucially on analysing the symmetries of the edge Lagrangian that describes a strictly 2d system. Hence it distinguishes between topological order that is allowed in strict 2d and ones that require a 3d bulk. For the case of $Z_2$ topological order we may, as usual,  take
\begin{equation}
K=
\left(
\begin{array}{cc}
 0   & 2  \\
  2 & 0
\end{array}
\right)
\label{KZ2}
\end{equation}
The corresponding Chern-Simons Lagrangian is simply
\begin{equation}
{\cal L} =  \frac{1}{\pi} a_e d a_m + \frac{1}{2\pi} A(\tau_e da_e + \tau_m da_m)
\end{equation}
Here $a_{1,2}$ are internal gauge fields and $A$ is an external `probe' gauge field. The charge vector $(\tau_e, \tau_m)$ has integer components.  Physically $da_{e,m}$ are $2\pi j_{e,m}$ of the $e$ and $m$ particles respectively. If both $e$ and $m$ carry global $U(1)$ charge $1/2$, then $\tau_e = \tau_m = 1$. As we already noted such a
charge assignment implies a non-zero Hall conductivity ad hence cannot describe a strict 2d system with time reversal invariance. Thus the surface topological order described above in Tables \ref{stou1rz2t} and \ref{stou1z2tp2} cannot occur in strict 2d systems. It remains to discuss the other two surface topological orders. In both (as well as the case in Table. \ref{stou1z2tp2} both $e$ and $m$ are Kramers doublets. We now show that this is not possible in any time reversal invariant strict 2d $Z_2$ topologically ordered state.

The $1+1$ dimensional edge theory corresponding to the Chern-Simons Lagrangian above is
\begin{equation}
{\cal L}_{edge} = \frac{1}{\pi} \partial_t \phi_e \partial_x \phi_m + .......
\end{equation}
with $a_{i1,2} = \partial_i \phi_{1,2}$. Demanding time reversal invariance of the edge Lagrangian we see immediately that the edge densities $\partial_x \phi_e, \partial_x \phi_m$ must transform with opposite signs.  However if  $e^{i\phi_e}$ creates one member of a Kramers doublet, it must transform as
\begin{equation}
e^{i\phi_e} \rightarrow ie^{-i\phi_e}
\end{equation}
under time reversal. Equivalently $\phi_e \rightarrow \phi_e + \frac{\pi}{2}$ so that the edge density $\partial_x \phi_e \rightarrow \partial_x \phi_e$. Thus if both $e$ and $m$ are Kramers doublets then both corresponding edge densities must be even under time reversal. But this is inconsistent with our deduction above from demanding time reversal invariance of the edge Lagrangian. We thus conclude that in strict 2d systems both $e$ and $m$ cannot be Kramers pairs in a time reversal invariant system. It is however alowed to happen at the surface of the 3d SPT phases described in this paper.
In passing we note that this precludes the possibility that strictly $2d$ spin models have gapped $Z_2$ topological phases where both non-trivial  bosonic quasiparticles carry spin-$1/2$ ({\em i.e} are spinons) while the fermionic quasiparticle carries no spin.

\end{document}